\documentstyle[epsfig,twoside]{article}

\catcode`\@=11
\long\def\@makefntext#1{
\protect\noindent \hbox to 3.2pt {\hskip-.9pt  
$^{{\eightrm\@thefnmark}}$\hfil}#1\hfill}		

\def\thefootnote{\fnsymbol{footnote}}
\def\@makefnmark{\hbox to 0pt{$^{\@thefnmark}$\hss}}	
	
\def\ps@myheadings{\let\@mkboth\@gobbletwo
\def\@oddhead{\hbox{}
\rightmark\hfil\eightrm\thepage}   
\def\@oddfoot{}\def\@evenhead{\eightrm\thepage\hfil
\leftmark\hbox{}}\def\@evenfoot{}
\def\sectionmark##1{}\def\subsectionmark##1{}}

\oddsidemargin=\evensidemargin
\renewcommand{\thefootnote}{\fnsymbol{footnote}}

\newcounter{sectionc}\newcounter{subsectionc}\newcounter{subsubsectionc}
\renewcommand{\section}[1] {\vspace{12pt}\addtocounter{sectionc}{1} 
\setcounter{subsectionc}{0}\setcounter{subsubsectionc}{0}\noindent 
	{\tenbf\thesectionc. #1}\par\vspace{5pt}}
\renewcommand{\subsection}[1] {\vspace{12pt}\addtocounter{subsectionc}{1} 
	\setcounter{subsubsectionc}{0}\noindent 
	{\bf\thesectionc.\thesubsectionc. {\kern1pt \bfit #1}}\par\vspace{5pt}}
\renewcommand{\subsubsection}[1] {\vspace{12pt}\addtocounter{subsubsectionc}{1}
	\noindent{\tenrm\thesectionc.\thesubsectionc.\thesubsubsectionc.
	{\kern1pt \tenit #1}}\par\vspace{5pt}}
\newcommand{\nonumsection}[1] {\vspace{12pt}\noindent{\tenbf #1}
	\par\vspace{5pt}}

\newcounter{appendixc}
\newcounter{subappendixc}[appendixc]
\newcounter{subsubappendixc}[subappendixc]
\renewcommand{\thesubappendixc}{\Alph{appendixc}.\arabic{subappendixc}}
\renewcommand{\thesubsubappendixc}
	{\Alph{appendixc}.\arabic{subappendixc}.\arabic{subsubappendixc}}

\renewcommand{\appendix}[1] {\vspace{12pt}
        \refstepcounter{appendixc}
        \setcounter{figure}{0}
        \setcounter{table}{0}
        \setcounter{lemma}{0}
        \setcounter{theorem}{0}
        \setcounter{corollary}{0}
        \setcounter{definition}{0}
        \setcounter{equation}{0}
        \renewcommand{\thefigure}{\Alph{appendixc}.\arabic{figure}}
        \renewcommand{\thetable}{\Alph{appendixc}.\arabic{table}}
        \renewcommand{\theappendixc}{\Alph{appendixc}}
        \renewcommand{\thelemma}{\Alph{appendixc}.\arabic{lemma}}
        \renewcommand{\thetheorem}{\Alph{appendixc}.\arabic{theorem}}
        \renewcommand{\thedefinition}{\Alph{appendixc}.\arabic{definition}}
        \renewcommand{\thecorollary}{\Alph{appendixc}.\arabic{corollary}}
        \renewcommand{\theequation}{\Alph{appendixc}.\arabic{equation}}
        \noindent{\tenbf Appendix \theappendixc #1}\par\vspace{5pt}}
\newcommand{\subappendix}[1] {\vspace{12pt}
        \refstepcounter{subappendixc}
        \noindent{\bf Appendix \thesubappendixc. {\kern1pt \bfit #1}}
	\par\vspace{5pt}}
\newcommand{\subsubappendix}[1] {\vspace{12pt}
        \refstepcounter{subsubappendixc}
        \noindent{\rm Appendix \thesubsubappendixc. {\kern1pt \tenit #1}}
	\par\vspace{5pt}}

\newcommand{\textlineskip}{\baselineskip=13pt}
\newcommand{\smalllineskip}{\baselineskip=10pt}

\def\eightcirc{
\begin{picture}(0,0)
\put(4.4,1.8){\circle{6.5}}
\end{picture}}
\def\eightcopyright{\eightcirc\kern2.7pt\hbox{\eightrm c}} 

\newcommand{\copyrightheading}[1]
	{\vspace*{-2.5cm}\smalllineskip{\flushleft
{\footnotesize Prepared March 8, 2000 for:\\ 
International Journal of Modern Physics E, #1}\\
{\footnotesize World Scientific Publishing Company}\\
	 }}

\def\abstracts#1#2#3{{
	\centering{\begin{minipage}{4.5in}\baselineskip=10pt\footnotesize
	\parindent=0pt #1\par 
	\parindent=15pt #2\par
	\parindent=15pt #3
	\end{minipage}}\par}} 

\renewenvironment{thebibliography}[1]
	{\frenchspacing
	 \ninerm\baselineskip=11pt
	 \begin{list}{\arabic{enumi}.}
        {\usecounter{enumi}\setlength{\parsep}{0pt}     
	 \setlength{\leftmargin 12.7pt}{\rightmargin 0pt} 
         \setlength{\itemsep}{0pt} \settowidth
	{\labelwidth}{#1.}\sloppy}}{\end{list}}

\newcounter{itemlistc}
\newcounter{romanlistc}
\newcounter{alphlistc}
\newcounter{arabiclistc}

\newenvironment{arabiclist}
	{\setcounter{arabiclistc}{0}
	 \begin{list}{\arabic{arabiclistc}}
	{\usecounter{arabiclistc}
	 \setlength{\parsep}{0pt}
	 \setlength{\itemsep}{0pt}}}{\end{list}}

\newcommand{\fcaption}[1]{
        \refstepcounter{figure}
        \setbox\@tempboxa = \hbox{\footnotesize Fig.~\thefigure. #1}
        \ifdim \wd\@tempboxa > 5in
           {\begin{center}
        \parbox{5in}{\footnotesize\smalllineskip Fig.~\thefigure. #1}
            \end{center}}
        \else
             {\begin{center}
             {\footnotesize Fig.~\thefigure. #1}
              \end{center}}
        \fi}

\newcommand{\tcaption}[1]{
        \refstepcounter{table}
        \setbox\@tempboxa = \hbox{\footnotesize Table~\thetable. #1}
        \ifdim \wd\@tempboxa > 5in
           {\begin{center}
        \parbox{5in}{\footnotesize\smalllineskip Table~\thetable. #1}
            \end{center}}
        \else
             {\begin{center}
             {\footnotesize Table~\thetable. #1}
              \end{center}}
        \fi}

\def\@citex[#1]#2{\if@filesw\immediate\write\@auxout
	{\string\citation{#2}}\fi
\def\@citea{}\@cite{\@for\@citeb:=#2\do
	{\@citea\def\@citea{,}\@ifundefined
	{b@\@citeb}{{\bf ?}\@warning
	{Citation `\@citeb' on page \thepage \space undefined}}
	{\csname b@\@citeb\endcsname}}}{#1}}

\newif\if@cghi
\def\cite{\@cghitrue\@ifnextchar [{\@tempswatrue
	\@citex}{\@tempswafalse\@citex[]}}
\def\citelow{\@cghifalse\@ifnextchar [{\@tempswatrue
	\@citex}{\@tempswafalse\@citex[]}}
\def\@cite#1#2{{$\null^{#1}$\if@tempswa\typeout
	{IJCGA warning: optional citation argument 
	ignored: `#2'} \fi}}

\def\pmb#1{\setbox0=\hbox{#1}
	\kern-.025em\copy0\kern-\wd0
	\kern.05em\copy0\kern-\wd0
	\kern-.025em\raise.0433em\box0}

\def\fnt#1#2{\footnotetext{\kern-.3em
	{$^{\mbox{\scriptsize #1}}$}{#2}}}

\def\fpage#1{\begingroup
\voffset=.3in
\thispagestyle{empty}\begin{table}[b]\centerline{\footnotesize #1}
	\end{table}\endgroup}

\def\runninghead#1#2{\pagestyle{myheadings}
\markboth{{\protect\footnotesize\it{\quad #1}}\hfill}
{\hfill{\protect\footnotesize\it{#2\quad}}}}
\font\tenrm=cmr10
\font\tenit=cmti10 
\font\tenbf=cmbx10
\font\bfit=cmbxti10 at 10pt
\font\ninerm=cmr9

\font\eightrm=cmr8

\def\qed{\hbox{${\vcenter{\vbox{			
   \hrule height 0.4pt\hbox{\vrule width 0.4pt height 6pt
   \kern5pt\vrule width 0.4pt}\hrule height 0.4pt}}}$}}

\renewcommand{\thefootnote}{\fnsymbol{footnote}}	

\def\bsc{{\sc a\kern-6.4pt\sc a\kern-6.4pt\sc a}}	
\def\bflatex{\bf L\kern-.30em\raise.3ex\hbox{\bsc}\kern-.14em 
T\kern-.1667em\lower.7ex\hbox{E}\kern-.125em X} 

\begin{document}
\hyphenation{stran-gen-ess had-rons had-roni-za-tion}

\runninghead{Jean Letessier and Johann Rafelski.
} 
{
Observing  Quark-Gluon Plasma with Strange Hadrons.}

\normalsize\textlineskip
\thispagestyle{empty}
\setcounter{page}{1}

\copyrightheading{}			

\vspace*{0.88truein}

\fpage{1}
\centerline{\bf OBSERVING QUARK-GLUON PLASMA}
\vspace*{0.035truein}
\centerline{\bf WITH STRANGE HADRONS}
\vspace*{0.37truein}
\centerline{\footnotesize JEAN LETESSIER}
\vspace*{0.015truein}
\centerline{\footnotesize\it Laboratoire de Physique Th\'eorique et Hautes 
Energies\footnote{LPTHE, Univ.\,Paris 6 et 7 is:
Unit\'e mixte de Recherche du CNRS, UMR7589.}}
\baselineskip=10pt
\centerline{\footnotesize\it Universit\'e Paris 7, 2 place Jussieu, F--75251 Cedex 05}
\vspace*{10pt}
\centerline{\footnotesize and}
\vspace*{10pt}
\centerline{\footnotesize JOHANN RAFELSKI\footnote{Supported 
by U.S. Department of Energy under grant  DE-FG03-95ER40937\,.}}
\vspace*{0.015truein}
\centerline{\footnotesize\it Department of Physics, 
University of Arizona, Tucson, AZ 85721}

\vspace*{0.21truein}
\abstracts{
 We review the  methods and results  obtained in an 
analysis of the experimental  heavy ion collision
research program at nuclear beam energy of 160--200$A$\ GeV. 
We study  strange, and more generally, hadronic 
particle production experimental data. We  discuss present
expectations concerning how these observables will perform at other
collision energies. 
We also present the dynamical theory of strangeness production
and apply it to show that it agrees with available experimental 
results. We describe strange hadron production from the
baryon-poor quark-gluon 
phase formed at much higher reaction energies, where the
abundance of strange baryons and antibaryons exceeds that of nonstrange 
baryons and antibaryons.%
\footnote{PACS: {12.38.Mh, 25.75.-q, 25.75.Dw, 25.75.Ld}}}{}{}


\vspace*{1pt}\textlineskip	
\setcounter{footnote}{0}      
\renewcommand{\thefootnote}{\alph{footnote}}
\section{Introduction and Overview}\label{introsec}
\subsection{Introduction}
\vspace*{-0.5pt}
\noindent
The study of highly excited and dense hadronic matter by means of 
ultra-relativistic nuclear collisions is a relatively novel area 
of research at the border between nuclear and particle physics.
As such it, is in a rapid experimental and theoretical evolution. The
primary goal of research in this field is the creation
and investigation of elementary (particle) matter under extreme 
density and temperature conditions.
The existence of a novel non-nuclear high temperature phase of 
elementary matter is an unavoidable consequence 
of the current knowledge about the strong nuclear interaction, 
rooted in the theory of strong interactions, 
the quantum field theory of quarks and gluons called
quantum chromodynamics (QCD).\cite{FGL73,Pol74}

Discovery and study of quark-gluon plasma (QGP), a `deconfined'
state  consisting of mobile, color-charged quarks and gluons, is the
objective of the relativistic heavy ion experimental research
program  underway at the Relativistic Heavy Ion Collider (RHIC) at 
Brookhaven National Laboratory (BNL), New York, and
at the Super Proton Synchrotron (SPS) 
accelerator at the European Organization for Nuclear Research 
(CERN), Geneva.\cite{QGP} 
In a recent half day long workshop, and in an accompanying
press release in February 2000
the CERN laboratory has formally announced that it views the collective 
evidence  available today in their seven  relativistic nuclear collision
experiments as being conclusive: 
``\ldots{\it  
A common assessment of the collected data leads us to conclude that 
we now have compelling evidence that a new state of matter has indeed 
been created, at energy densities which had never been reached over 
appreciable volumes in laboratory experiments before and which 
exceed by more than a factor 20 that of normal nuclear matter. 
The new state of matter found in heavy ion collisions at the SPS 
features many of the characteristics of the theoretically predicted 
quark-gluon plasma.
}\ldots''.\cite{CERN}

This research program has  been developed over the past two decades in order to 
study the properties of elementary matter at conditions similar to those seen 
in the very early universe 30$\mu s$ after the big-bang, 
before the temperature decreases to about 
$T=150$\,MeV$\approx 1.7\cdot 10^{12}$\,K.
It has been many times argued that it is possible to achieve a laboratory 
recreation of this condition in a small-bang relativistic nuclear collision. 
The question is at what collision energy the transition to 
a color deconfined QGP phase first 
occurs. Early suggestion has been that this could
occur at an intrinsic available 
energy per participating nucleon 
as low  as  4 and 8 times the nucleon mass, 
corresponding to the range  30 and 120$A$\,GeV per
nucleon beam energy in fixed target experiments.\cite{RD83}

The conditions in the early universe and those created in nuclear
collision experiments differ somewhat: whereas the primordial
quark-gluon plasma survived for about 30 $\mu$s in the big bang,
the comparable conditions in nuclear collisions are not expected
to last for more than $10^{-22}$\,s due to the rapid explosion of
the hot matter ``fireball''. Moreover, in the matter created in heavy ion
collisions quarks are expected to outnumber antiquarks noticeably
due to the baryon content of the colliding nuclei, whereas the net
relative excess of the quarks over antiquarks in the universe was
less than $10^{-9}$. 

Numerical simulations of QCD suggest that 
the nature of the transformation between 
the hadronic and quark-gluon phases can change drastically 
as the values of the parameters of the theory 
are varied.\cite{LGT1,LGT2,LGT3} 
Recent analytical studies of the phase properties of 
QCD have supported the conclusion that the dependence on the 
net baryon density (baryochemical potential) is especially
interesting.\cite{SRS98,Raj99} Perhaps the most fundamentally important
observable in this context is the latent heat associated with 
the breaking of color bonds among quarks, leading to the
deconfinement of quarks. An experimental determination of this 
quantity and its dependence on beam energy would be of great 
scientific interest. 

Strange particle signatures for the formation and evolution
of the deconfined quark-gluon phase of elementary matter 
form a significant cornerstone of experimental QGP discovery. 
This subject has been  developed quite intensely for the 
past 20 years.\cite{firstS,RM82,KMR86,ER91,Raf91,acta96} 
The enabling difference in physics between confined and 
deconfined matter concerning strange particle  
signatures is rather simple:\\
$\bullet$ In the QGP phase the particle density is high enough 
and the strange flavor production energy threshold low enough to
assure that a high abundance of strangeness 
can actually be produced on the time scale 
available\cite{RM82,KMR86,acta96,MS86,BCD95}
while in the confined matter phase this
has been shown not to be the case,\cite{KMR86,KR85}
as long as one wants to remain consistent with other 
experimental results.\\ 
$\bullet$ Population at, and even in excess, of chemical equilibrium of 
hadron phase space occupancies occurs only when the
entropy rich QGP phase disintegrates rapidly and explosively 
into hadrons.\cite{KMR86,Raf91,acta96,Let97}

There are several important and when viewed together, uniquely
QGP characteristic predictions regarding strangeness, expected 
to occur should deconfinement set in. 
Specifically, the three pillars on which the QGP hypothesis stands
when seen by means of strangeness flavor observables are: \\
\indent 1) matter-antimatter symmetry as seen for directly 
emitted strange baryon and antibaryon 
particles in the $m_\bot$-spectral shape and strange quark fugacity;\\
\indent 2) (multi)strange baryon and antibaryon enhancement 
increasing with strangeness content;\\
\indent 3) enhancement of the (specific)
strangeness flavor yield per reaction participant 
(baryon),  by a factor 1.5--3  at SPS conditions, the value
depending on what is used as baseline, and if one looks alone at the 
central rapidity region, where this effect is strongest, or
considers the global strangeness yield, including the kinematic
domains of projectile and target fragmentation.

{\it All three predictions have recently been confirmed at 
the current SPS energy range 158 GeV per nucleon
for  Lead (Pb), and some also for  200 GeV per nucleon
Sulphur (S) induced reactions.}\\ 
Examples and stepping stones are in particular:\\
\indent 1) The WA97 collaboration  reported 
a detailed study of transverse mass strange baryon and 
antibaryon spectra which show a highly unusual
symmetry between strange baryon and antibaryon sector.\cite{Ant00}\\
\indent 2) 
A detailed analysis of Pb--Pb results by the WA97 collaboration 
has demonstrated, comparing p--p,  p--A with A--A results, 
a strong enhancement in the pertinent (multi) 
strange  baryon and  antibaryon yields, 
increasing with strangeness content.\cite{Ant99,WA97gen,WA97a,Lie99}
The results of the NA49 collaboration are consistent 
with these findings.\cite{App98} The WA85 collaboration also
finds an enhancement of multi-strange baryons and
antibaryons, increasing with strangeness content in S-induced 
reactions.\cite{WA85}\\
\indent 3a) 
Strangeness enhancement at mid-rapidity has been observed in 
S-induced reactions by the experiments NA35,\cite{Alb94} 
WA85 and WA94,\cite{Eva99} and NA44.\cite{NA44S} 
In the larger Pb-Pb reaction system strangeness
enhancements are reported by WA97,\cite{Ant99,WA97gen,WA97b} 
 NA49,\cite{NA49T} and NA44.\cite{NA44}
Results of the experiment NA52 suggest further that the 
onset of strangeness enhancement occurs rather suddenly 
as the centrality of the collisions and thus the size of
participating matter rises above baryon number 
$B$ = 40--50\,.\cite{Kabana}\\
\indent 3b)
Global Strangeness enhancement has been observed both in S-induced and
in Pb-Pb reactions by the NA35,\cite{Alb94} and NA49,\cite{NA49T} 
experiments.

In this article, we also 
rely in many aspects of this discussion indirectly
on other experimental results of NA35 and NA49 collaborations,
which offer a global view on particle production pattern 
considering the large kinematic acceptance.\cite{Alb95,App99}
We will not discuss or use here other experimental discoveries which 
have contributed to the CERN announcement, which are not
related to strangeness, such as $J/\Psi$ suppression,
dilepton and direct photon production. Readers interested in these
topics  should consult the list of experimental results.\cite{CERN}

In view of many often intricate but, when analyzed, 
convincing experimental findings about strange particle production, 
the purpose of this article is to present 
a comprehensive and selfconsistent  view on the 
understanding of the evidence comprised in strange hadron production
for the formation of quark-gluon plasma at CERN, 
and to discuss resulting expectations how this 
observable will perform at RHIC.
We  address in this review:\\
\phantom{ii}i) 
in {section\bf~2} the status of a  analysis  of the 
experimental data obtained at SPS,\cite{actaZ99}\\
\phantom{i}ii) 
in {section\bf~3} the implications of these results for the understanding of 
the dense\\ 
\phantom{iii) }phase formed in  these reactions,\\
iii) in {section\bf~4} an adaptation of the dynamical theory of 
strangeness production\\
\phantom{iii) } in QGP to RHIC conditions,\\
iv) in {section\bf~5} an application of these results to obtain
predictions for hyperon\\
\phantom{iii) }yields from QGP at RHIC,\\
\phantom{i}v) in {section\bf~6} highlights of the results
presented here and we draw our conclusions. 

\subsection{Overview}
\vspace*{-0.5pt}
\noindent
We introduce in {section\bf~2},
the Fermi-2000 model,\cite{acta96,LRa99} a straightforward 
elaboration of the original Fermi proposal,\cite{Fer50} 
that final state strongly interacting particles are
produced with a probability commensurate to the size of the 
accessible phase space. In this approach, the hadron phase space is 
characterized to the required degree of accuracy by six parameters
which have a clear physical meaning and can be  in 
 computed, and/or rather easily understood qualitatively, when their
values have been determined by an analysis of the experimental data. 
The physical picture underlying the use of the statistical Fermi
model in the 21st century is the sudden, explosive disintegration
of a high temperature hadronic matter fireball, 
apparently consisting of deconfined quark-gluon matter. 
Since its proposal 50 years ago, the Fermi model has been subject to 
considerable scrutiny and adaptation, with  
Hagedorn's `boiling' hadronic matter being the most important
stepping stone.\cite{HAG} The 
following were the relevant recent steps in the development of
the statistical particle production description
required to analyze the strange particle experimental results: 
\begin{enumerate}
\item 
Considering the interest and considerable theoretical effort  
vested in understanding  strange quark production 
mechanisms and the study of chemical equilibration 
processes,\cite{RM82,KMR86} it was a  natural  
refinement to introduce  an
expression of chemical non-equilibrium in the number of 
strange quark pairs, $\gamma_s\ne 1$,\cite{Raf91} into
 the analysis of strange hadrons.
\item 
In the analysis of the entropy content in S--W/Pb 
interactions,\cite{Let93} we found entropy
excess related to excess of meson abundance. This prompted 
us to explore possible nonequilibrium yield of mesons compared 
to that of baryons.\cite{Let95}
\item 
Since the dense hadron fireball is subject to explosive 
disintegration, final state hadrons emerge from  rapidly 
outward moving volume cells. In order to describe quantitatively 
spectra of hadronic particles, and their yields in restricted domains
of phase space, such `collective' matter flow motion 
needs to be modeled.\cite{Let95,Hei92,Hei93}
\item
Chemical (particle abundance changing) processes occurring at the time of 
hadronization do not generally lead to an equilibrium chemical yield 
of light  quarks, and the chemical non-equilibrium is more
pronounced if hadronization is a sudden process on the time scale of 
chemical quark equilibration. While this effect has been well accepted for 
strange quarks, as these need to be produced in microscopic processes, the 
need to introduce the  light quark pair abundance parameter, $\gamma_q\ne 1$,
was recognized rather late,\cite{LRa99,LRPb99} and is not yet widely accepted. 
\end{enumerate}
A data analysis we perform, allowing for all these effects, does not simply yield a 
set of `best' parameters; rather:\\
\phantom{ ii}i) 
it offers a complete characterization of the phase space of hadrons and
its occupancy, 
allowing one to extrapolate reliably the particle yields into kinematic
domains not accessible at present; \\
\phantom{ i}ii) 
it allows one to study and understand the magnitude of 
model  parameters, so that we can safely  to extrapolate their
values to other reaction conditions, {\it e.g.}, from SPS to RHIC as will
be done here;\\
\phantom{ }iii) 
it allows one to evaluate the physical properties of 
the phase space characterized by
these  parameters, which provides very precise information about the 
 physical properties of the particle source. This in turn leads to 
the understanding of the nature of the dense fireball created
in the heavy ion collision.

In order to pursue these aims, we need to reach considerable 
precision in the description of experimental results. 

Our approach and objectives elaborate significantly on the now 
commonly accepted  observation that  all hadronic particles produced  
in strong interaction processes{\it qualitatively} satisfy 
statistical model predictions, as discovered and
discussed in great detail by Rolf 
Hagedorn more than 30 years  ago.\cite{HAG}
 The fact  that the statistical model indeed `works' 
 does not cease to amaze and impress.\cite{BHS99,CR99} 
At times this even provokes the hypothesis  that `thermal'
abundances of hadrons could arise in some mysterious 
and unknown way,\cite{CS93} and thus one could proceed to predict 
`thermal' yields of hadrons, apparently believing 
that the statistical `thermal'
model substitutes for  conventional particle production mechanisms.
The recent proposal by Bialas to consider the fluctuation of string tension 
is indeed suggesting how such an 
explanation could arise in p--p reactions.\cite{Bia99} 

However, for reactions of large nuclei, 
the statistical description implies and exploits the 
result of repeated occurrence of microscopic collision  reactions,
and the associated approach to an equilibrium distribution shape, and
{\it independently}, also approach to chemical particle abundance
equilibrium. This is independent of the above described 
possibility that in  elementary
reaction systems  equilibration is possibly 
a consequence of microscopic properties 
of strong interactions. Our discussion of A--A reactions thus aims at 
an improvement of the statistical description beyond the `thermal' model,
such that we can deal with standard deviation errors 
comparing theory and experiment, as is common
in the field of particle and nuclear physics. 

The way we set up the  Fermi-2000 model represents the microscopic 
processes that are occurring, and of course there are limits
to this description. For example, 
primary  high energy initial interactions can produce heavy quark 
flavor which could not arise from the generally
softer interactions occurring in the kinetically equilibrated 
system (an example is expected production of charm at 
RHIC\cite{CRH,Bas00}).  More generally, an acceptable 
failure of a statistical description is the one which
under-predicts the yield of rarely produced particles. 
For this reason, it is necessary to scrutinize the validity of the 
statistical description, at least for the most rarely produced 
hadrons. 

Pertinent  results of a complete
 analysis of the Pb--Pb system are presented in 
{section\bf~3}. 
We describe abundances and spectra of hadronic particles 
observed  by both the wide acceptance 
NA49-experiment and the central rapidity  (multi)strange (anti)baryon  
WA97-experiment.
Our method of analysis shows that  both these families of results 
obtained with widely different
methods are consistent and it allows us to reach the required precision in
their description. This objective could be reached only after 
we have introduced light quark chemical non-equilibrium and allowed that
the all strange $\Omega(sss)$ and $\overline\Omega(\bar s\bar s\bar s)$
hadrons are enhanced beyond their statistical phase space yield, a point 
we will discuss in greater detail in { subsection\bf~3.4} below.
These developments occurred after  
the last comprehensive review of the subject 
 appeared,\cite{Sol97} and after the extensive SPS-Pb-beam experimental 
results  became available. The introduction of
light quark chemical  nonequilibrium has had a significant 
impact on the determination
of the physical properties of the hadron emitting source,
as  it allows for a considerable reduction of the 
chemical freeze-out temperature: specifically  we have determined that 
$T_f=143\pm5$\,MeV, we have also
included here an estimate of the systematic error, for 
the temperature at which practically all strange hadrons are formed.
This result is nearly 30\,MeV below values that one might 
infer otherwise in a qualitative study of the statistical 
hadron yields.\cite{BHS99,CR99,Bec98}

This relatively low freeze-out temperature is consistent with 
the result that the chemical freeze-out parameters
determine correctly the shape of hadron $m_\bot$-spectra, which
suggests that after
the deconfined QGP source has dissociated the resulting hadrons
are practically free-streaming and thus that thermal and chemical
freeze-out do not differ much if at all.
 This particle production scheme is called 
{\it sudden hadronization}.\cite{KMR86,Raf91}
In terms of a microscopic model it occurs when 
hadronic particles are  produced either~in:\\ 
\indent a) an evaporation process from a hot 
expanding  surface, or\\ 
\indent b) a sudden global hadronization process of exploding,
possibly  super-cooled  deconfined matter.

Experimental evidence supporting the picture of
sudden QGP hadronization  is most directly 
derived  from  the baryon-antibaryon transverse
mass $m_\bot$-spectral  symmetry.  Another piece of
evidence for a sudden hadronization is  the 
chemical overabundance  of light quark pairs in hadronization
and the associated maximization of entropy density in hadron phase 
space as will be discussed in subsection {\bf 2.3}. Not to be forgotten is
the original observation about experimental data 
that has led to the data interpretation in terms of the sudden
hadronization model\cite{Raf91}: the strange quark fugacity  as measured 
by emitted strange hadrons implies a source with freely moving 
strange $s$ and antistrange $\bar s$ quarks such that 
$\langle s -\bar s\rangle=0$, a point we will discuss in subsection~{\bf 2.2}.

In subsection~{\bf 3.3}, we study in depth the phenomenon of strangeness 
enhancement and show that the rather precise analysis results we obtain are
in excellent agreement with the theoretically computed strangeness yields,
assuming formation of the QGP phase. We also show that, 
in the case of Pb--Pb, the explosive disintegration of the dense QGP fireball 
leads to an overpopulation of the strangeness phase space abundance, and show that 
theoretical results are again in good agreement with the results of the 
data analysis. The initial temperatures for this agreement to occur are 
in the range $260<T_{\rm ch}<320$, which values we obtain from models
of collision dynamics. 

In   { section\bf~4},
we develop the theoretical method which leads to the finding of 
overpopulated strangeness phase space discussed in {subsection\bf~3.3}.
This, at a first sight  surprising result,  occurs
due to early freeze-out of strangeness abundance in a rapid 
explosive evolution of the QGP fireball.  We find a similar
non-equilibrium result,  in section{\bf~5}, for 
 RHIC condition in presence of 
transverse expansion which increases the speed at which QGP dilutes.
We explore the dynamics of the phase space
occupancy, rather than particle density, which allows us to
eliminate much of the dependence on the
dynamical flow effects by incorporating in the dynamics considered
the hypothesis of entropy  conserving  matter flow and evolution.

Exploiting the experience with SPS data analysis, we are
able to consider, in  {section\bf~5},
strange particle production   at RHIC.
We obtain an unexpected  particle abundance pattern: during the 
hadronization of the baryon-poor RHIC-QGP phase there is considerable
advantage for strangeness flavor to stick to baryons and antibaryons.
This can be easily understood realizing that production
 of strange (anti)baryons   is favored over production of kaons
by the energy balance, {\it i.e.}: $E(\Lambda+\pi)<E(\mbox{N+K})$. 
Moreover, at RHIC  there are a high number of strange quarks per baryon
available, and in this strangeness bath just a few (anti)baryons will
manage to emerge without strangeness content. We thus expect and find 
in detailed study in { section\bf~5}
that hyperon production dominates  baryon production, {\it i.e.},
most baryons and antibaryons produced  will be strange. We consider  this 
result to be a unique consequence of the sudden QGP hadronization 
scenario observed at SPS and hope and expect that hyperon dominance
should, when observed at  RHIC, be generally accepted as proof of
formation of the deconfined phase of nuclear matter. This phenomenon 
shows how much more pronounced will be the physics of
strangeness in QGP
at RHIC, as compared to the ten times lower SPS energy range.

\section{Contemporary Fermi Model of Hadron Production}
\label{fermisec}
\subsection{Phase space and parameters}
\noindent
The relative number of  final state hadronic particles
freezing out from, {\it e.g.}, a thermal quark-gluon source is obtained
noting that the  fugacity $f_i$ of the $i$-th emitted  composite  
hadronic particle containing $k$-components is derived 
from fugacities $\lambda_k$ and phase space occupancies $\gamma_k$:
\begin{equation}\label{abund}
N_i\propto e^{-E_i/T_f}f_i=e^{-E_i/T_f}\prod_{k\in i}\gamma_k\lambda_k.
\end{equation}
We study chemical properties of light quarks $u,d$ jointly, 
denoting these by a single index $q$ and also
consider chemical properties of strange quarks $s$\,. 
Thus as seen in Eq.\,(\ref{abund}), we study particle 
production in terms of five statistical parameters. In addition
there is at least one matter flow velocity
parameter. The six parameters which 
characterize the accessible phase space of hadronic particles
made of light quarks `q' and strange quarks `s'
and their natural values, assuming a QGP source, are 
in turn:\\
\indent 1) $\lambda_{s}$: The value of strange quark 
fugacity $\lambda_{s}$ cab be obtained  
from the  requirement that strangeness  balances,
$\langle N_{s}-N_{\bar s}\rangle=0\,,$
which for a source in which all $s,\bar s$ quarks are unbound and thus 
have symmetric phase space, implies $\lambda_{s}=1$\,.
 However, the Coulomb distortion
of the strange quark phase space plays an important role in the
understanding of this constraint for Pb--Pb collisions,\cite{LRPbC99} 
leading to the Coulomb-deformed value $\lambda_{s}\simeq 1.1$\,,
as discussed in next subsection.\\
\indent 2) $\gamma_{s}$: The strange quark phase space occupancy 
$\gamma_{s}$ can be computed, and will be studied in 
this review in detail within the framework 
of kinetic theory.\cite{RM82,acta96} For a rapidly
expanding system the production processes will lead to an
over-saturated  phase space with $\gamma_{s}>1$\,. 
The difference between the two different types of chemical parameters
$\lambda_i$ and $\gamma_i$ is that the phase space
occupancy  factor $\gamma_{i}$ regulates the number of pairs of flavor `$i$', 
and hence applies in the same manner to particles and antiparticles, while
fugacity $\lambda_i$ applies only to particles, while $\lambda_i^{-1}$ is
the antiparticle fugacity. \\
\indent 3) $\lambda_{q}$: The light quark fugacity $\lambda_{q}$, 
or equivalently, the baryochemical potential:
\begin{eqnarray}\label{muB}
\mu_B=3\,T_{f}\ln \lambda_{q},
\end{eqnarray}
regulate the baryon density of the fireball and hadron  freeze out. 
This density can vary 
dependent on the energy and size of colliding nuclei, and  thus 
the value of $\lambda_{q}$ is not easily predicted. 
However, since we know the energy per baryon content in 
the incoming nuclei, if we assume that the deposition of the 
baryon number and energy (`stopping') is similar, 
we know the energy per baryon content in
the fireball.\cite{acta96} This qualitative knowledge can be used in  a study of
equations of state applicable to the dense fireball to establish a constraint. \\
\indent 4) $\gamma_{q}$: The equilibrium phase space occupancy of light quarks 
$\gamma_{q}$ is expected to  significantly exceed unity 
to accommodate the excess  entropy content in the 
plasma phase.\cite{Let93,Let95}
There is an upper limit:
\begin{equation}\label{gammaqc}
\gamma_{q}<\gamma_{q}^c\equiv e^{m_\pi/2T}\,,
\end{equation}
which arises if all pions  produced are simultaneously present, forming a 
Bose gas. We will address this effect in subsection {\bf 2.3}.\\
\indent 5) $T_{f}$: The freeze-out temperature $T_{f}$ is  expected to be 
not much different from the Hagedorn temperature 
$T_H\simeq 160$\,MeV,\cite{HAG} which characterized particle production in 
proton-proton reactions. \\
\indent 6) $v_c$: The  collective  expansion velocity $v_c$ is expected 
to remain below  the relativistic sound velocity\cite{acta96}: 
\begin{equation}
v_c\le 1/\sqrt{3}.
\end{equation}
When the source emitting the free streaming 
particles is undergoing local collective flow motion, 
spectra of particles emitted are described by replacing
the Boltzmann factor in Eq.\,(\ref{abund}) by:
\begin{eqnarray}\label{abundflow}
e^{-E_i/T}&\to& \frac1{2\pi}\int d\Omega_v
  \gamma_c(1-\vec v_{\rm c}\cdot \vec p_i/E_i)
  e^{-{{\gamma_cE_i}\over T}
    \left(1-\vec v_{\rm c}\cdot \vec p_i/E_i\right)},\nonumber\\
\gamma_c&=&\frac1{\sqrt{1-\vec v_{\rm c}^{\,2}}}\,,
\end{eqnarray}
a result which can  be intuitively obtained by a Lorentz
transformation between an observer on the surface of
the fireball, and one at rest in the laboratory frame. Formal
derivation of this and more elaborated results 
requires a considerably more precise framework.\cite{Hei92}

The resulting yields of final state hadronic particles are most 
conveniently characterized taking the Laplace transform of the 
accessible phase space. This approach generates a  function which
in its mathematical properties is identical to 
the partition function. For example
for the open strangeness sector we find, for the case $v_{\rm c}=0$:
\begin{eqnarray}
\ln{\cal Z}_{s} = { {V T^3} \over {2\pi^2} }
\hspace{-0.3cm}&&\left\{
(\lambda_{s} \lambda_{q}^{-1} + \lambda_{s}^{-1} \lambda_{q}) 
      \gamma_{s} \gamma_{q} F_K 
+(\lambda_{s} \lambda_{q}^{2} + \lambda_{s}^{-1} \lambda_{q}^{-2}) 
\gamma_{s}\gamma_{q}^{2}  F_Y \right.\nonumber \\ 
&&\hspace{0.3cm}\left.+ (\lambda_{s}^2 \lambda_{q} +
\lambda_{s}^{-2} \lambda_{q}^{-1}) 
\gamma_{s}^2\gamma_{q}  F_\Xi + (\lambda_{s}^{3} + \lambda_{s}^{-3})
\gamma_{s}^3  F_\Omega\right\}\,.
\label{4a}
\end{eqnarray}
The integrated momentum phase space 
factors $F_i$ for kaons $i=K$, single strange 
hyperons  $i=Y$, doubly strange cascades 
$i=\Xi$ and triply strange omegas $i=\Omega$ 
are:
\begin{equation}
F_i=\sum_j g_{i_j} W(m_{i_j}/T)\,,\qquad W(x)=x^2K_2(x)\,.
\label{FSTR}
\end{equation}
$g_{i_j}$ is the statistical degeneracy of each contributing hadron 
resonance `$j$' of the kind `$i$' in the  $\sum_j$, which comprise all 
known strange hadron resonances.  $K_2$  is the modified Bessel function which
arises from the relativistic phase space integral of the
thermal particle distribution $f(\vec p)\propto e^{-\sqrt{m^2+p^2}/T}$.
It is important to keep in mind that:\\
a) Eq.\,(\ref{4a}) does not require formation of a phase comprising a
gas of hadrons,  but is not inconsistent with such a step in evolution 
of the fireball; in that sense  it is not a partition function, but just 
a look-alike object arising from the  Laplace transform of the accessible 
phase space, and \\
b) the final particle abundances measured in an experiment 
are obtained after all unstable hadronic resonances `$j$'
are allowed to disintegrate, contributing to the yields of
stable hadrons;\\
c) the unnormalized particle multiplicities arising are obtained 
differentiating Eq.\,(\ref{4a}) with respect to particle 
fugacity. The relative particle yields are simply 
given by ratios of corresponding chemical factors, weighted with
the size of the momentum phase space accepted by the experiment.
For particles showing the same spectral shape comparison of
normalization
of $m_\bot$ spectra suffices, {\it e.g.}, Ref.\,\cite{Raf91}:
\begin{equation}\label{ex1}
\left.\frac{\Xi^-(dss)}{\Lambda(dds)}\right\vert_{m_\bot}=
\frac{g_\Xi\gamma_d\gamma_s^2\lambda_d\lambda_s^2}
{g_\Lambda\gamma_d^2\gamma_s\lambda_d^2\lambda_s}\,.
\end{equation}
$g_i$ are the spin statistical factors of the states considered.
Similarly:
\begin{equation}\label{ex2}
\left.\frac{\overline{\Xi^-(dss)}}
  {\overline{\Lambda(dds)}}\right\vert_{m_\bot}=
\frac{g_\Xi\gamma_d\gamma_s^2\lambda_d^{-1}\lambda_s^{-2}}
{g_\Lambda\gamma_d^2\gamma_s\lambda_d^{-2}\lambda_s^{-1}}\,.
\end{equation}
When acceptance is limited to central rapidity, and significant 
flow is present considerable effort must be made to introduce 
appropriate phase space weights. \\
d) In some experimental data it is important to distinguish 
the two light quark flavors as is in fact the case in the two above examples.
This can be incorporated considering how the average light quark fugacity 
varies between both light quark species,\cite{Raf91} and assuming that
the phase space occupancies are equal.

We consider, for SPS energy range,
 the radial flow model, which is without doubt the  simplest of
the reasonable and expected matter flow cases possible, in
view of the behavior of global observables seen in 
these experiments. As the results below
show, this suffices to  assess the impact of 
collective flow on the data analysis originally developed 
to be as little as possible sensitive to collective matter
flow, even when particle yields in highly restricted 
regions of  $m_\bot,\,y$ are considered. The collective source flow
can completely change the shape of momentum distribution of particle 
produced, though of course it  leaves  unchanged
the total particle yield, which is the integral sum 
of particle  multiplicity over the entire phase space of the flow 
spectrum. However,  particles of different mass 
experience differing flow effects when $m_\bot,\,y$ acceptance cuts are
present. Moreover, particles can freeze-out at 
slightly different conditions. In order  to  limit the influence of the 
practically unknown collective flow structure on particle yields in 
limited domains of the accessible phase space, we study compatible 
particle ratios: these are yield ratios obtained in a restricted domain of
$m_\bot,\,y$, for particles of similar mass and believed to have 
a similar interaction strength with the matter background.  

We now will address in turn two special topics, which slightly contradict
expectations, and thus require more attention. Firstly, we 
 review the the properties of the strange quark fugacity $\lambda_s$\,, 
which is sensitive to the possible 
asymmetry between strange and antistrange quarks in the source. The 
importance of this parameter is that it potentially helps distinguish 
the confined from deconfined phase: while in the baryon-rich 
confined phase the requirement
of strangeness conservation implies that $\lambda_s>1$\,, in the deconfined
phase the symmetry between phase space of strange and antistrange quarks
implies $\lambda_s\simeq 1$\,. Following this, we address in more 
detail case of pions,\cite{LTR00} which is exceptional since
we will be considering  rather large values of $\gamma_{q}>1.5$\,.
As we shall see, the pion gas emerging from the QGP phase
is strongly influenced  by Bose correlation effects; in fact it is
close to satisfying the Bose condensation condition.

\subsection{Coulomb force}
It has been  recognized for a long time that the Coulomb force can be
of considerable importance  in the study of relativistic heavy ion collisions.
It plays an important role in the HBT analysis of the structure 
of the particle source.\cite{Pratt}  We show
that the analysis of chemical properties at freeze-out is also subject
to this perturbing force, and in consideration of the precision reached in
the study  of particle ratios, one has to keep this effect in mind. 

We consider a Fermi  gas of strange and antistrange 
quarks, allowing that the Coulomb potential $V$ established by the excess 
charge of the colliding nuclei distorts significantly the phase space. 
Within a relativistic Thomas-Fermi phase space
occupancy model,\cite{MR75} and 
allowing for finite temperature in QGP we have\cite{LRPbC99}:
\begin{equation}\label{Nsls}
\langle N_s-N_{\bar s}\rangle =\!\!
 \int\limits_{R_{\rm f}}\! g_s\frac{d^3rd^3p}{(2\pi)^3}\!\left[
 \frac1{1+\gamma_s^{-1}\lambda_s^{-1}e^{(E(p)-\frac13 V(r))/T}}-
 \frac1{1+\gamma_s^{-1}\lambda_se^{(E(p)+\frac13 V(r))/T}}\right],
\end{equation}
which clearly cannot vanish for $V\ne 0$ in the limit $\lambda_s\to1$.
In Eq.\,(\ref{Nsls}) the subscript ${R_{\rm f}}$ on the spatial integral 
reminds us that only the classically
allowed region within the fireball is covered in the integration over the 
level density; $E=\sqrt{m^2+\vec p^{\,2}}$, and for a uniform charge distribution
within a radius $R_{\rm f}$ of charge $Z_{\rm f}$:
\begin{equation}
V=\left\{
\begin{array}{ll}\displaystyle
-\frac32 \frac{Z_{\rm f}e^2}{R_{\rm f}}
      \left[1-\frac13\left(\frac r{R_{\rm f}}\right)^2\right]\,,
          & \mbox{for}\quad r<R_{\rm f}\,;
  \\\displaystyle
-\frac{Z_{\rm f}e^2}{r}\,,& \mbox{for} \quad r>R_{\rm f}\,.
\end{array}
\right.
\end{equation}

One obtains a rather precise result for the range of parameters of interest
to us  using the Boltzmann approximation:
\begin{equation}\label{balance}
\langle N_s-N_{\bar s}\rangle =
\gamma_s\left\{\int g_s\frac{d^3p}{(2\pi)^3}e^{-E/T}\right\}
\int_{R_{\rm f}} d^3r\left[\lambda_s e^{\frac V{3T}}
 - \lambda_s^{-1} e^{-\frac V{3T}}\right]\,.
\end{equation}
The Boltzmann limit allows us also to verify and confirm the signs: the Coulomb
potential is negative for the negatively charged $s$-quarks with
the magnitude  of the charge, $1/3$, made explicit in the potential
terms in all expressions  above. We thus have\cite{LRPbC99}:
\begin{equation}\label{tilams}\label{lamQ}
\tilde\lambda_{s}\equiv \lambda_{s} \lambda_{\rm Q}^{1/3}=1\,,\qquad
\lambda_{\rm Q}\equiv
\frac{\int_{R_{\rm f}} d^3r e^{\frac V{T}} } {\int_{R_{\rm f}} d^3r}\,.
\end{equation}
 $\lambda_{\rm Q}<1$  expresses the Coulomb deformation of 
strange quark phase space. $\lambda_{\rm Q}$ is not a fugacity that 
can be adjusted to satisfy a chemical condition,
since consideration of $\lambda_i,\ i=u,d,s$, exhausts all available
chemical balance conditions for the abundances of hadronic particles. 
The subscript ${R_{f}}$, in Eq.\,(\ref{lamQ}), reminds us 
that the classically
allowed region within the dense matter fireball is included in
the integration over the  level density.
 Choosing $R_{\rm f}=8$\,fm, $T=140$\,MeV,
$m_{s}=200$\,MeV, noting that the value of $\gamma_{s}$ 
is practically irrelevant
as this factor cancels in Boltzmann approximation, see Eq.\,(\ref{balance}),
we find for $Z_{\rm f}=150$ that the value 
$\lambda_{s}=1.10$ corresponds to $R_{\rm f}=7.9$\,fm.
The Coulomb effect is thus  relevant in central Pb--Pb interactions, 
while for S--Au/W/Pb reactions,  similar analysis leads to a value 
$\lambda_{s}=1.01$, little different from the value $\lambda_{s}=1$
expected in the absence of the Coulomb phase space deformation. 
Another way to understand the
varying importance of the Coulomb effect 
is to note that while the Coulomb potential acquires in the Pb--Pb case
a magnitude comparable to the quark chemical potential, it remains 
small on this scale for S--Au/W/Pb reactions.

\subsection{Super-dense pion gas and chemical non-equilibrium}
\noindent
For pions composed of a light quark-antiquark pair, the chemical fugacity 
is $\gamma_{q}^2$, see Eq.\,(\ref{abund}). 
Thus the pion momentum space distribution 
has the Bose shape:
\begin{equation}\label{piBos}
f_\pi(E)=\frac{1}{\gamma_{q}^{-2}e^{E_\pi/T}-1}\,,\qquad E_\pi=\sqrt{m_\pi^2+p^2}\,.
\end{equation}
The range of values for $\gamma_{q}$ is bounded from above by the Bose
singularity. When $\gamma_{q}\to \gamma_{q}^c$, Eq.\,(\ref{gammaqc}),
the lowest energy state (in the continuum limit with $p\to 0$)
will acquire macroscopic  occupation and a pion condensate is formed. 
Formation of such a condensate `consumes' energy without
consuming entropy of the primordial high entropy QGP phase. 
On the other hand, as we shall see presently, when $\gamma_{q}\to \gamma_{q}^c$
the entropy content of the pion gas initially grows!
Thus while the development, directly from the QGP phase, of a pion 
condensate is not likely,  the sudden hadronization of entropy rich QGP
should lead to the limiting value $\gamma_{q}\to \gamma_{q}^c$, in
order to more efficiently connect the entropy rich deconfined and 
the confined phases. An interesting feature of such a mechanism 
of phase transition is that the chemical non-equilibrium reduces
and potentially eliminates any discontinuity in the phase transition, 
which thus, in the experiment, will appear more like a phase transformation
without critical fluctuations, even if theory implies a 1st order
phase transition for statistical equilibrium system.

\begin{figure}[tb]
\vspace*{-1.6cm}
\centerline{
\psfig{width=8.5cm,figure=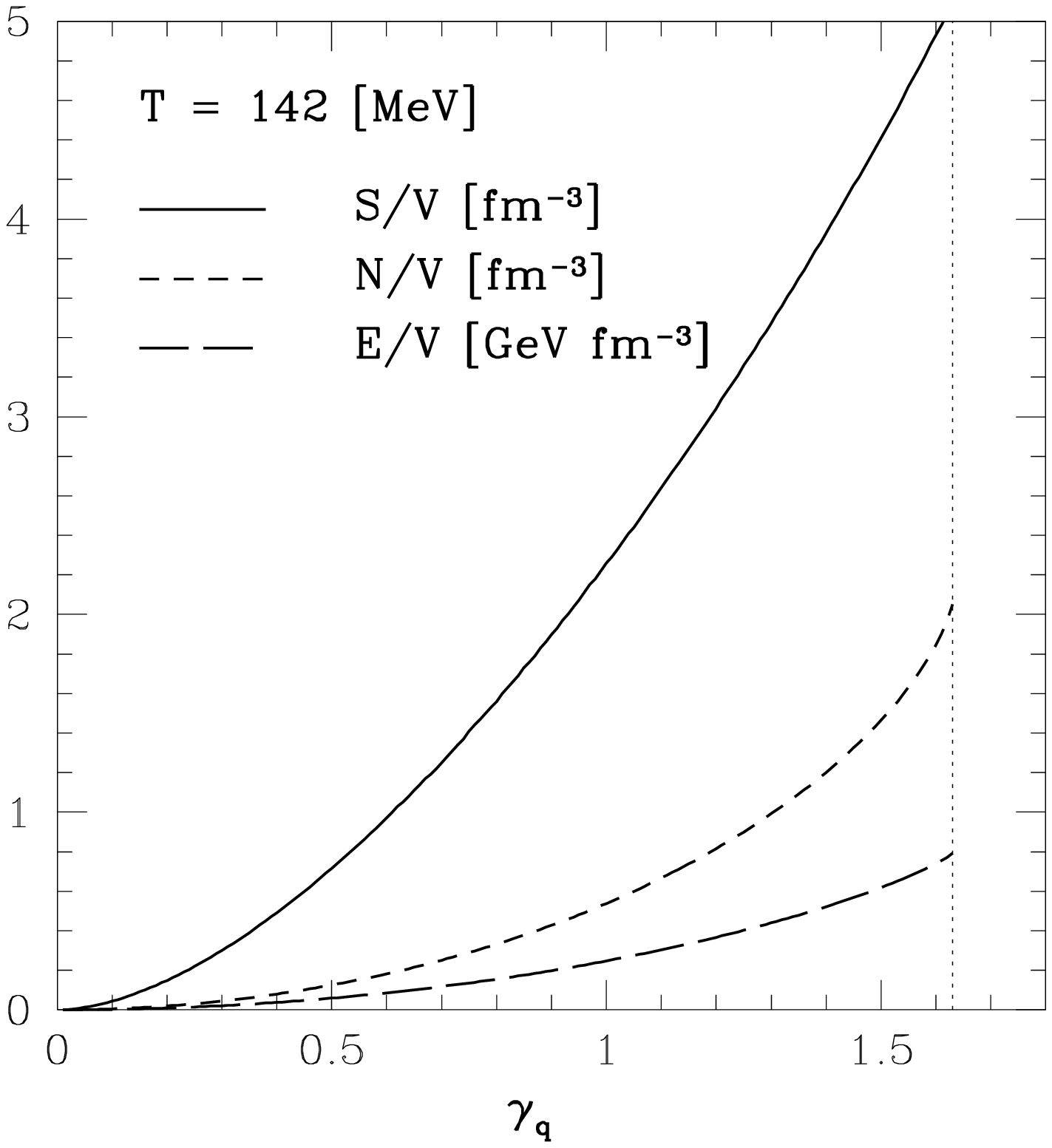}}
\vspace*{-2cm}
\fcaption{ 
Dependence of pion gas properties $N/V$-particle, $E/V$-energy and 
$S/V$-entropy  density, as function of  $\gamma_{q}$ at $T=142$\,MeV. 
\label{abssne}} 
\end{figure}

In Fig.\,\ref{abssne}, we show the physical properties 
of a pion gas as function of 
$\gamma_{q}$ for a gas temperature $T=142$\,MeV.\cite{LTR00} We see
(solid line) that a large range of entropy density can be accommodated 
by varying the parameter $\gamma_{q}$. It is important to remember
that in the hadronization of a quark-gluon phase it is relatively 
easy to accommodate energy density, simply by producing a few heavy hadrons.
However, such particles being in fact non-relativistic at the 
temperature considered, are not effective carriers of pressure
and entropy. However, as we see now in Fig.\,\ref{abssne}, 
the super-dense pion gas is just the missing element 
to allow a  rapid hadronization  process, since 
the entropy density is nearly twice as high at $\gamma_{q}\simeq 
\gamma_{q}^c$ than at $\gamma_q=1$. Without this phenomenon 
one has to introduce a mechanism that allows the parameter $VT^3$ 
to grow, thus expanding either the volume $V$ due to formation of so called
mixed phase or invoking rise of $T$ in so called reheating. 

\begin{figure}[t]
\vspace*{-1.3cm}
\centerline{ 
\psfig{width=6.8cm,figure=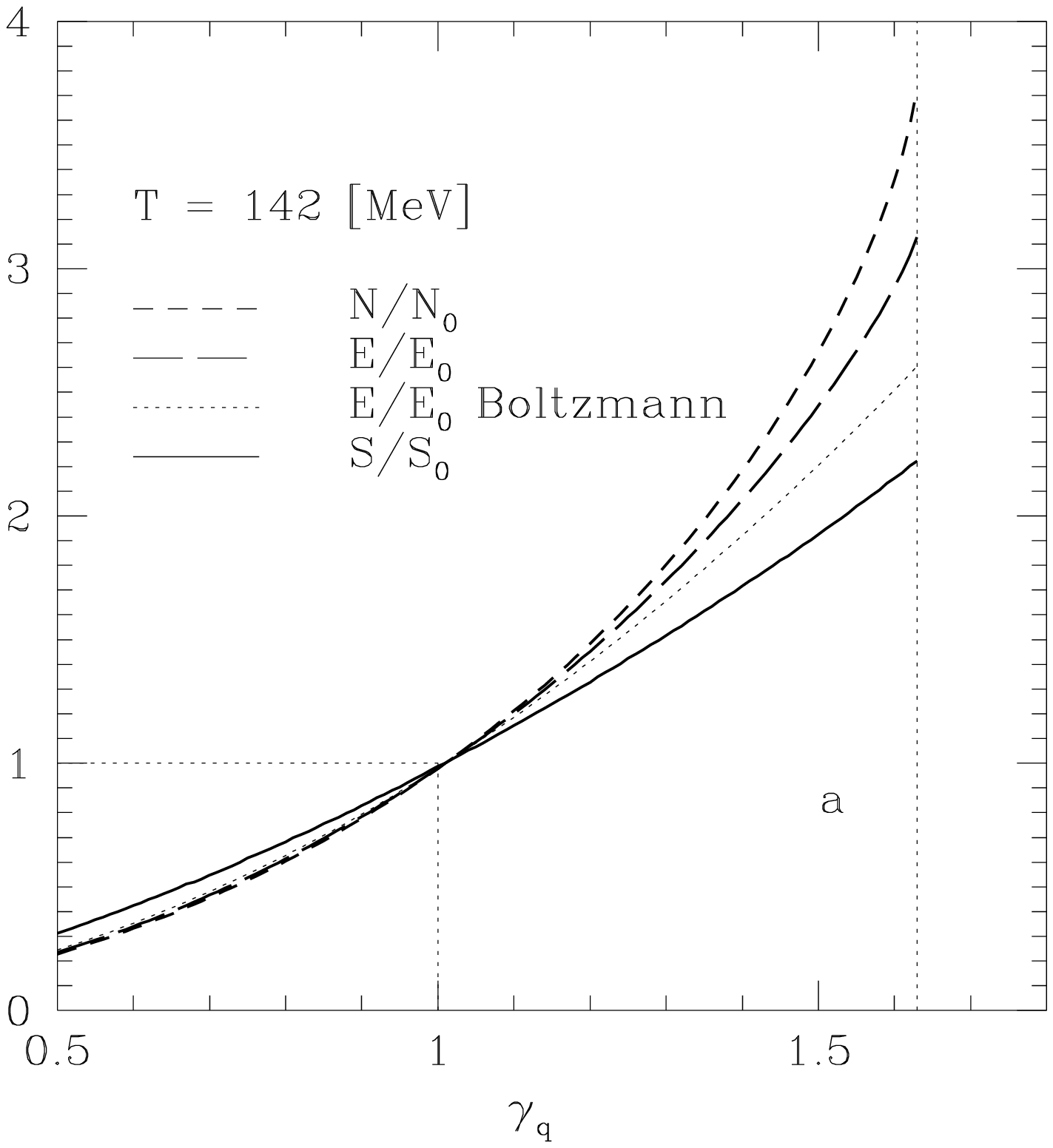}\hspace*{-0.5cm}
\psfig{width=6.8cm,figure=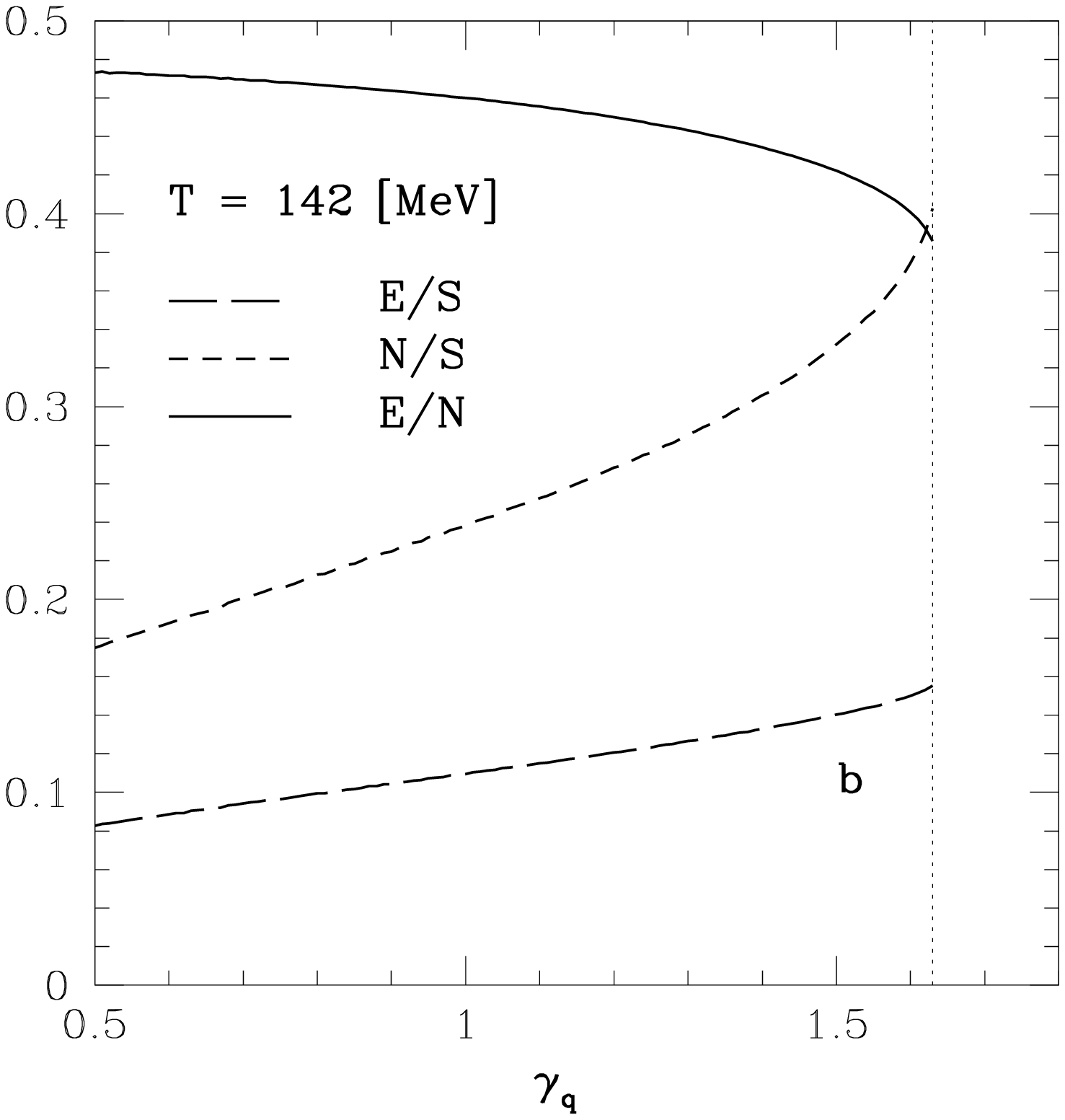}}
\vspace*{-1.5cm}
\fcaption{ 
Dependence of pion gas properties ($N$-particle, $E$-energy and 
$S$-entropy)  density 
as function of  $\gamma_{q}$ for $T=142$\,MeV. 
{\bf a)} ratios relative to equilibrium value $\gamma_q=1$; 
{\bf b)} relative ratios, thus $E/N$, $N/S$ and $E/S$.
\label{ratiosne}} 
\end{figure}

The specific properties of the super-dense 
pion gas are shown  Fig.\,\ref{ratiosne}. In  Fig.\,\ref{ratiosne}a,
we relate the properties to the chemical equilibrium value $\gamma_{q}=1$
and we also show that the Boltzmann approximation is not qualitatively wrong,
as long as $\gamma_{q}< \gamma_{q}^c$\,. In  Fig.\,\ref{ratiosne}b,
we see the relative change
in energy per pion,  (inverse of) entropy per pion, and
energy per entropy. Interestingly, we note that the entropy per pion 
drops as $\gamma_{q}$ increases, and at the condensation 
point $\gamma_{q}=\gamma_{q}^c$, we can add pions without increase in entropy.
We further note  that a hadronizing gas will consume, at higher $\gamma_q$, less
energy per particle, and that the energy per entropy is nearly
constant. 

It is important to remember that if the hadronization process were
adiabatic, allowing a full equilibrium relaxation, naturally 
${\gamma_i\to1}$ would arise: as is implicitly well known  
 the value ${\gamma_i\to1}$ maximizes the  entropy 
for a particle gas at fixed total energy, 
corresponding to the chemical equilibrium.\cite{entro} 
This result is easily found considering Boltzmann 
pion gas, and recalling in some detail the definition of  entropy,
since the standard equilibrium expressions do not apply:
\begin{eqnarray}\label{entro1}
{S}_{\rm B,F}=&&\hspace{-0.6cm}
 \int\!\frac{d^3\!p\, d^3\!x}{(2\pi\hbar)^3}\,
   \left[\pm(1\pm f(x,p))\ln(1\pm f(x,p))
-               f(x,p)\ln f(x,p)\right]\,,\\
\to&&\hspace{-0.6cm}\label{entro2}
 V\int\!\frac{d^3\!p }{(2\pi\hbar)^3}f(p)\ln[e/f(p)]\,,
\end{eqnarray}
where aside of `B,F' (Bose, Fermi) also the Boltzmann limit for a homogeneous 
spatial distribution is shown explicitly. 

Evaluating in Boltzmann limit
the particle number and  energy, we find that
the factor $\gamma_q^2$ becomes a normalization factor which describes the
average occupancy of the phase space relative to the equilibrium
value, and for entropy, we also find a logarithmic term:
\begin{eqnarray}
N=&&\hspace{-0.6cm}
     \gamma_q^2 N|_{\rm eq}\to aV\gamma_q^2 T^3\,,\label{NBol}\\
E=&&\hspace{-0.6cm}
     \gamma_q^2 E|_{\rm eq}\to 3aV\gamma_q^2 T^4\,,\label{EBol}\\
S=&&\hspace{-0.6cm}
     \gamma_q^2 S|_{\rm eq} +\ln \left(
\gamma_q^{-2}\right)\gamma_q^2\, N|_{\rm eq} 
\to 4aV\gamma_q^2 T^3+\ln \left( \gamma_q^{-2} \right)
     aV\gamma_q^2 T^3\,.\label{SBol}
\end{eqnarray}
For massless pions,  $a=g/\pi^2$, with pion  degeneracy $g=3$\,. 
Setting $E=$\,Const., we eliminate $T$ and find that 
the entropy as function of $\gamma_q$ varies according to:
\begin{equation}
S\vert_{E=\mbox{Const.}}\propto \gamma_q^{1\over 2}(4-\ln\gamma_q^2)\,,
\end{equation}
which has a very weak maximum at $\gamma_q=1$;
note that at $\gamma=1.4$, the entropy is at 98.3\% of the value at
$\gamma=1$.  At this point it is important to realize that the 
chemical equilibrium is much better defined for the more familiar case of a
fixed temperature bath (and not an isolated fixed energy fireball discussed
above):  consider the free energy
$\cal F$ of the  (non-interacting) relativistic  gas at fixed
temperature $T$.  Since ${\cal F}=E-TS$, we combine 
Eqs.\,(\ref{NBol}--\ref{SBol}) and obtain in the Boltzmann limit,
\begin{equation}
{\cal F}^l=-aVT^4\gamma_q^2\left[1+\ln\left(\gamma_q^{-2}\right)\right]
\,,
\end{equation}
which  has a minimum for chemical equilibrium value $\gamma_q=1$. However, one now
finds that a change by factor 1.4 in $\gamma_q$, at fixed $T$ leads
to a change by 35\% in the value of the free energy and even a greater
change in entropy. 
 
We conclude that, in adiabatic condition, the fireball would evolve to 
the maximum entropy equilibrium case $\gamma_i=1, i=q,s$, the gain in 
entropy for an isolated system is in this limit very minimal. Thus 
for a system with rapidly evolving volume, 
we will in general find more effective
 paths to increase entropy, other than the establishment of 
the absolute chemical equilibrium. Hence the values we report 
$\gamma_i\ne1, i=q,s$, are consistent with the 
present day understanding of explosive  evolution of 
the hadronic matter fireball.

In a systematic study of the relevance of different physical
parameters, the  chemical non-equilibrium at hadron freeze-out
has been  shown to be a required ingredient in order to 
arrive at a precise interpretation of the experimental 
results on particle ratios $R^j$ obtained at  CERN. 
This is best seen considering the 
results for the statistical parameters obtained for the 
S--Au/W/Pb collisions,\cite{LRa99} and the associated  total 
statistical error, 
\begin{equation}\label{chi2}
\chi^2_{\rm T}\equiv\frac{\sum_j({R_{\rm th}^j-R_{\rm exp}^j})^2}{
({{\Delta R _{\rm exp}^j}})^2}\,,
\end{equation}
which are presented in table \ref{fitsw}. 
We clearly see the gain in physical significance
that is accomplished as chemical non-equilibrium is allowed for
 by releasing the fixed value $\gamma_i=1$, first for strange 
and  next, light quarks.  We also observe that allowing for $\lambda_{s}\ne 1$  
does not lead to an improvement in statistical significance, since 
the data is compatible with 
this  value expected for the deconfined QGP. Similar systematic 
study has also been completed  for the Pb--Pb system,\cite{LRPb99}  
reconfirming the need to use  $\gamma_{i}\ne 1$ in the data analysis.

The  errors in the results shown in table \ref{fitsw}, and in results 
that follow below, are one standard  deviation errors arising
from the propagation of the experimental measurement error. However, 
these errors are meaningful only when the theoretical model describes 
the data well, as is the case for last entry line in table \ref{fitsw}
when we allow light quark chemical nonequilibrium, $\gamma_{q}\ne 1$\,.

\begin{table}[tb]
\tcaption{\label{fitsw}
Statistical parameters obtained from  fits of S--Au/W/Pb data
without enforcing strangeness conservation, and not considering 
flow effect --- only compatible particle ratios were considered.
Asterisk ($^*$) means a fixed (input) value. See text for an 
explanation of result errors.}
\begin{center}
\begin{tabular}{ccccc|c}\\
\hline
$T_{\rm f}$ [MeV]& $\lambda_{ q}$&$\lambda_{s}$&
$\gamma_{s}$&$\gamma_{q}$& $\chi^2_{\rm T}/$dof \\
\hline
  145 $\pm$ 3
                 & 1.52 $\pm$ 0.02
                 &   1$^*$
                 &   1$^*$
                 &   1$^*$
                 &   17  \\
 144 $\pm$ 2
                 & 1.52 $\pm$ 0.02
                 &   0.97 $\pm$ 0.02
                 &   1$^*$
                 &   1$^*$
                 &   18  \\
 147 $\pm$ 2
                 & 1.48 $\pm$ 0.02
                 &   1.01 $\pm$ 0.02
                 &   0.62 $\pm$ 0.02
                 &   1$^*$
                 &   2.4  \\
 144 $\pm$ 3
                 &  1.49 $\pm$ 0.02
                 &  1.00 $\pm$ 0.02
                 &  0.73 $\pm$ 0.02
                 &  1.22 $\pm$ 0.06
                 &  0.90  \\
\hline
\end{tabular}
\end{center}
\end{table}

\section{Strange Hadron Data Analysis}
\label{spsec}
\subsection{Particle yields}
\noindent
The available compatible particle yield ratios (excluding 
presently $\Omega$ and $\overline\Omega$, see { subsection\bf~3.4})
 are listed  in table~\ref{resultpb2}, top section from the
experiment WA97, for $p_\bot>0.7$ GeV, within a narrow
$\Delta y=0.5$ central rapidity window. Further below 
are shown  results from the large  acceptance experiment NA49, 
extrapolated by the collaboration to full $4\pi$ phase space coverage. 
We first fit 11 experimental results shown in table~\ref{resultpb2},
and then turn to include also the $m_\bot$-slope  into this consideration,
and thus have 12 data points.  The total error $\chi^2_{\rm T}$
for the  four result columns is shown at the bottom of this table
along with the number of data points `$N$', parameters `$p$' 
used and  (algebraic) redundancies `$r$' connecting the 
experimental results. For $r\ne 0$ it is more appropriate 
to quote the total  $\chi^2_{\rm T}$, with a initial qualitative
statistical relevance condition  $\chi^2_{\rm T}/(N-p)<1$.

The first theoretical columns refer to results without
collective velocity $v_c$ (subscript $0$) the three other 
were allowing for $v_c$ (subscript $v_c$). In column three, superscript `sb' 
means that $\lambda_{s}$ is fixed by strangeness balance and, in column 
four, superscript `sc' means that $\gamma_{q}=\gamma_{q}^c=e^{m_\pi/2T_f}$, 
that is $\gamma_{q}$ is fixed by its upper limit, the pion condensation point. 
All results shown account for slightly higher value of the
ratio  $h^-/B$ recently reported\cite{App99}; $B$ is here the number of baryon
participants and  $h^-=\pi^-+K^-+\bar p$ is the yield of stable negative
hadrons comprising as indicated pions, kaons and antiprotons.

First we note that all columns in table \ref{resultpb2} represent physically 
acceptable result for the Pb--Pb collision system: \\
\indent a) presence of collective flow (three last columns) 
leads to very similar compatible particle ratios, 
even though improvement of $\chi_{\rm T}$
occurs when $v_c\ne 0$ is allowed for;\\
\indent b) the highest confidence result is found just when the light quark 
phase space occupancy assumes a 
value at below the pion condensation point;\\
\indent c) strangeness conservation (enforced in second last column) 
is naturally present, enforcing it does not change in any way 
the results for particle multiplicities. 

\begin{table}[tb]
\tcaption{\label{resultpb2}
WA97 (top) and NA49 (bottom)  Pb--Pb 158$A$ GeV particle ratios
compared with theoretical results, see text for explanation.}
\begin{center}
\begin{tabular}{lcl|l|lll}\\
\hline
 Ratios & Ref. &  Exp. Data                           &Pb$|_0$& Pb$|_v$ & Pb$|_v^{\rm sb}$ & Pb$|_v^{\rm sc}$  \\
\hline
${\Xi}/{\Lambda}$ & \cite{Kra98} &0.099 $\pm$ 0.008                    &  0.104  & 0.103  & 0.105 & 0.103\\
${\overline{\Xi}}/{\bar\Lambda}$ & \cite{Kra98} &0.203 $\pm$ 0.024     &  0.214  & 0.208  & 0.209 & 0.206\\
${\bar\Lambda}/{\Lambda}$  & \cite{Kra98} &0.124 $\pm$ 0.013           &  0.124  & 0.125  & 0.124 & 0.125\\
${\overline{\Xi}}/{\Xi}$  & \cite{Kra98} &0.255 $\pm$ 0.025            &  0.256  & 0.252  & 0.248 & 0.251\\
\hline
$(\Xi+\bar{\Xi})\over(\Lambda+\bar{\Lambda})$& \cite{Ody97}  &0.13 $\pm$ 0.03
                                                                      &  0.126  & 0.122  & 0.124 & 0.122\\
${K^0_{s}}/\phi$   & \cite{Puh98}  & 11.9 $\pm$ 1.5\ \             &  14.2   & 13.3   & 13.0  & 13.4 \\
${K^+}/{K^-}$         & \cite{Bor97}         &  1.80$\pm$ 0.10         &  1.80   & 1.82   & 1.78  & 1.83 \\
$p/{\bar p}$     & \cite{Ody98}              &18.1 $\pm$4.\ \ \ \      &  17.3   & 16.7   & 16.6  & 16.6 \\
${\bar\Lambda}/{\bar p}$     & \cite{Roh97}  & 3. $\pm$ 1.             &  2.68   & 2.11   & 2.11  & 2.11 \\
${K^0_{s}}$/B       & \cite{Jon96}       & 0.183 $\pm$ 0.027       &  0.181  & 0.181  & 0.163 & 0.188\\
${h^-}$/B                 & \cite{App99}     & 1.97 $\pm $ 0.1\ \      &  1.96   & 1.97   & 1.97  & 1.96 \\
\hline
 & $\chi^2_{\rm T}$     &                                             &  3.6    & 2.5    & 3.2   & 2.6  \\
 &  $ N;p;r$     &                                                    & 11;5;2  & 12;6;2 & 12;5;2& 12;5;2\\
\hline\\
\end{tabular}
\end{center}
\vskip -0.8cm
\end{table}

Allowing  radial flow not only improves  the capability to describe the 
data, but it allows us to study  $m_\bot$ particle spectra, which
 offer another independent measure 
of flow, and confirm  the value
of $v_c$ --- when considering $v_c$ along with  $ T_{\bot}$,
the inverse slope of the $m_\bot$ spectra,
 we have one parameter and several spectral inverse slopes of
particles considered. However, we will  in the first instance 
assume that we have just one additional  data point 
and we proceeded as follows: for a given pair of values
 $T_{f}$ and $v_{\rm c}$ we evaluate the resulting
$m_\bot$ particle spectrum and analyze it using the spectral shape
and kinematic cuts employed by the experimental groups.
Once we find  values of $T_{\rm f}$
and $v_{\rm c}$, we  study again the inverse slopes of 
individual  particle spectra and obtain an acceptable
agreement with the experimental $T_{\bot}^j$ as 
shown in left section of table~\ref{Tetrange2}\,.
We have considered in the same framework the
S-induced reactions, and the right section of 
table~\ref{Tetrange2} shows also a  good
agreement with the WA85 experimental data.\cite{WA85slopes}

\begin{table}[tb]
\tcaption{\label{Tetrange2}
Experimental and theoretical $m_\bot$ spectra inverse slopes $T_{\rm th}$.
Left Pb--Pb results  from experiment WA97\protect\cite{Lie99,Ant00};
right  S--W  results from  WA85.\protect\cite{WA85slopes}}
\begin{center}
\begin{tabular}{lcc|cc}\\
\hline
               & $T_{\bot}^{\rm Pb}$\,[MeV]&$T_{\rm th}^{\rm Pb}$\,[MeV]&$T_{\bot}^{\rm S}$\ [MeV]&
                                                                   $T_{\rm th}^{\rm S}$\ [MeV]\\
\hline
$T^{{\rm K}^0}\ \ $         & 230 $\pm$ \phantom{1}2&  241& 219 $\pm$  \phantom{1}5 &  215\\
$T^\Lambda$                 & 289 $\pm$ \phantom{1}3&  280& 233 $\pm$  \phantom{1}3 & 236\\
$T^{\overline\Lambda}$      & 287 $\pm$ \phantom{1}4&  280& 232 $\pm$  \phantom{1}7 & 236\\
$T^\Xi$                     & 286 $\pm$ \phantom{1}9&  298& 244 $\pm$  12& 246\\
$T^{\overline\Xi}$          & 284 $\pm$           17&  298& 238 $\pm$  16& 246\\
\hline\\
\end{tabular}
\end{center}
\vskip -0.8cm
\end{table}

We have updated the experimental Pb--Pb results shown in table~\ref{Tetrange2}
with the current high precision results.\cite{Ant00} However,  the theoretical 
results shown were obtained
 earlier for slightly different
results with larger error bars, and we hope to reevaluate these results
in the near future. To model these slopes theoretically, one needs to
remember that the vast majority of $\Lambda$ and $\overline\Lambda$ is
a decay product of $\Sigma^0$ and $\overline{\Sigma^0}$, $\Lambda^*$ and $\overline{\Lambda^*}$ 
and $\Xi$ and $\overline\Xi$. Consequently,  given the precision of the (inverse) 
slopes presented,  in order to model  the 
$\Lambda$ and $\overline\Lambda$ spectra one will need to consider the effect of 
hadron cascading, which introduces  uncertainty arising from a
dependence on unmeasured yields. However, given the current availability  
of quite precise $\Xi$ and $\overline\Xi$ slopes, and the fact that these particles
are rarely decay products of other hadronic resonances,
we will in future use these slopes as the spectral data point 
input in the data analysis studies. As result, we anticipate a 
slight reduction in the  collective velocity within 
the errors shown below. 

\subsection{Chemical freeze-out properties}
\noindent
The six  parameters ($T_f, v_c, \lambda_q, \lambda_s, \gamma_q, \gamma_s$)
describing the  particle abundances
are shown in the top section of table~\ref{fitqpbs}. We
also show in the last column  the best result for S-induced
reactions, where the target has been W/Au/Pb.\cite{LRa99}  
 All results shown  in 
table~\ref{fitqpbs} have convincing statistical confidence level.
For the S-induced reactions the number 
of redundancies $r$ shown in the heading of the table~\ref{fitqpbs}
is large, since the same data comprising
different kinematic cuts has been included in the analysis. 

Within error, the freeze-out 
temperature $T_{\rm f}\simeq 143\pm3$\,MeV, seen in table~\ref{fitqpbs},
is the same for both the S- and Pb-induced reactions,
even though the chemical phase space occupancies differ greatly. 
Such a behavior is expected in view of the similarity of the energy content 
in the collision in both reaction systems, but greatly differing 
collision geometry. We find that the variation in the 
shape of  particle $m_\bot$-spectra  is fully explained by a change in
the collective velocity, which  rises from
$v_c^{\rm S}=0.49\pm0.02$ to 
$v_c^{\rm Pb}=0.54\pm0.04\simeq 1/\sqrt{3}=0.577$.
The value of light quark fugacity $\lambda_{q}$ implies that baryochemical
potential is 
$\mu_B^{\rm Pb}=203\pm5 >\mu_B^{\rm S}=178\pm5$\,MeV.
As in  S-induced reactions where $\lambda_{s}=1$,
now  in Pb-induced reactions, a value $\lambda_{s}^{\rm Pb}\simeq 1.1$
characteristic for a source of freely movable  strange quarks with
balancing strangeness, {\it i.e.}, $\tilde\lambda_{s}=1$, is obtained, 
see Eq.\,(\ref{lamQ}). 

\begin{table}[tb]
\tcaption{\label{fitqpbs}
In heading, we present the total quadratic relative error 
$\chi^2_{\rm T}$, number of data points $N$, parameters $p$ and 
redundancies $r$;  in the upper section: statistical model parameters
which best describe the experimental results for
Pb--Pb data, and in last column for S--Au/W/Pb data presented in 
Ref.\,\protect\cite{LRa99}.
Bottom section: specific energy, entropy, anti-strangeness, net strangeness
 of  the full hadron phase space characterized by these
statistical parameters. In column two, we fix $\lambda_{s}$ by requirement of 
strangeness conservation, and in column three, we fix $\gamma_{q}$ at
the pion condensation point $\gamma_{q}=\gamma_{q}^c$.}
\vspace{-0.2cm}\begin{center}
\begin{tabular}{lccc|c}\\
\hline
                       & Pb$|_v$            & Pb$|_v^{\rm sb}$ & Pb$|_v^{\rm sc}$        & S$|_v$ \\
$\chi^2_{\rm T};\ N;p;r$&2.5;\ 12;\,6;\,2   & 3.2;\ 12;\,5;\,2 & 2.6;\ 12;\,5;\,2        &  6.2;\ 16;\,6;\,6 \\
\hline
$T_{f}$ [MeV]          &    142 $\pm$ 3     &  144 $\pm$ 2     &  142 $\pm$ 2            &  144 $\pm$ 2 \\
$v_c$                  &   0.54 $\pm$ 0.04  & 0.54 $\pm$ 0.025 & 0.54 $\pm$ 0.025        &   0.49 $\pm$ 0.02\\
$\lambda_{q}$          &   1.61 $\pm$ 0.02  & 1.605 $\pm$ 0.025& 1.615 $\pm$ 0.025       &   1.51 $\pm$ 0.02 \\
$\lambda_{s}$          &   1.09 $\pm$ 0.02  & 1.10$^*$         & 1.09 $\pm$ 0.02         & 1.00 $\pm$ 0.02   \\
$\gamma_{q}$           &   1.7 $\pm$ 0.5    & 1.8$\pm$ 0.2   &${\gamma_{q}^c}^*=e^{m_\pi/2T_f}$&   1.41 $\pm$ 0.08 \\
$\gamma_{s}/\gamma_{q}$&   0.79 $\pm$ 0.05  & 0.80 $\pm$ 0.05  & 0.79 $\pm$ 0.05         &  0.69 $\pm$ 0.03  \\
\hline
$E_{f}/B$              &   7.8 $\pm$ 0.5    & 7.7 $\pm$ 0.5    & 7.8 $\pm$ 0.5           &  8.2 $\pm$ 0.5    \\
$S_{f}/B$              &    42 $\pm$ 3      & 41 $\pm$ 3       & 43 $\pm$ 3              &   44 $\pm$ 3     \\
${s}_{f}/B$            &  0.69 $\pm$ 0.04   & 0.67 $\pm$ 0.05  & 0.70 $\pm$ 0.05         &   0.73 $\pm$ 0.05 \\
$({\bar s}_f-s_f)/B\ \ $   &  0.03 $\pm$ 0.04   & 0$^*$            &  0.04 $\pm$ 0.05        &    0.17 $\pm$ 0.05\\
\hline\\
\end{tabular}
\end{center}
\vskip -0.8cm
\end{table}

Further evidence for low chemical freeze-out temperature 
is contained in the 
$m_\bot$-particle spectra considered in subsection {\bf 3.1}. 
Our approach offers a natural understanding of 
the  equality  of the $m_\bot$-slopes  of the
strange baryons and antibaryons considered
which arises because within the sudden hadronization
model both these particles emerge  free-streaming from QGP.
In the hadron based microscopic 
simulations this behavior of  $m_\bot$-slopes of baryons
and antibaryons arises from fine-tuning of the 
particle-dependent freeze-out times.\cite{HSX98}
On the other hand, in such a microscopic study one
finds in view of the small reaction cross sections that 
$\Omega$ and  $\overline\Omega$ could freeze out somewhat
sooner than the other hadrons, and thus would have a softer
spectrum as also confirmed in direct hadronization
simulations.\cite{Bas99} We will return to this point 
just below.
The reader should  keep in mind  that since we find a rather low
chemical freeze-out temperature, and can explain the $m_\bot$ spectra
well based on this value, the implied kinetic (collision) freeze-out 
temperature must be rather  similar to the chemical freeze-out. 

In the bottom section of table~\ref{fitqpbs}, we also see the
energy and entropy content per baryon.
The  energy  per baryon seen in the emitted hadrons is nearly
equal to the available specific energy
of the collision  (8.6 GeV for Pb--Pb, 8.8--9 GeV for S--Au/W/Pb).
This implies that the fraction of energy deposited in the central
fireball  must be nearly (within 10\%) 
the same as the fraction of baryon number.
The small reduction of the specific entropy in Pb--Pb compared to
the lighter S--Au/W/Pb system maybe driven by the greater baryon
stopping in the larger system, also seen in the smaller energy per
baryon content. Both collision systems freeze out at 
the same energy per unit of entropy, 
$$E/S=0.185\,\mbox{GeV}\,. $$
There is a loose relation of this universality in the 
chemical freeze-out condition with the suggestion made
recently that particle freeze-out occurs at a fixed energy per baryon for
all physical systems,\cite{CR98} considering that the entropy content is related to
particle multiplicity. The overall high specific entropy content we find
agrees well with the entropy content evaluation we made  
earlier for the S--Pb case.\cite{Let93} 
The high entropy content is observed in the final hadron state in 
terms of enhanced pion yield. Thus the ratio  of $K^+/\pi^+$
is combines these two effects and is not a good indicator of new 
physics, even though this relatively simple observable continues to
attract attention.\cite{DO00} It would have been more useful if 
systematic studies of strangeness production and enhancement 
were to offer as result of their analysis the strangeness 
yields per participating baryon number.

The large values of $\gamma_{q}>1$, seen in table~\ref{fitqpbs}, imply  
as discussed earlier that there is
phase space over-abundance of light quarks, which receives contribution from,
{\it e.g.}, gluon fragmentation at QGP breakup.  
$\gamma_{q}$ assumes in the data 
analysis a value near to  where pions 
could begin to  condense,\cite{LRprl99} Eq.\,(\ref{gammaqc}). 
This result is consistent with the expectations for hadronization of 
an entropy rich quark gluon plasma, as we discussed above in 
{subsection\bf~2.3}.
We found by studying the ratio $h^-/B$
separately from other experimental results
that the value of $\gamma_{q}\simeq\gamma_{q}^c$ is fixed
consistently and independently both, by the negative hadron ($h^-$),
and the strange hadron yields. The unphysical  range 
$\gamma_{q}>\gamma_{q}^c\simeq 1.63$ can  arise 
(see column Pb$\vert_v^{\mbox sb}$)  since, up to this
point, we had used only a first quantum (Bose/Fermi) 
correction. However, when  Bose distribution for 
pions is implemented, which requires  the 
constraint $\gamma_{q}\le\gamma_{q}^c$, 
we obtain  practically the same results, as shown
in the third column of table~\ref{fitqpbs}.

\subsection{Strangeness enhancement}
\noindent
We show, in the bottom section of table~\ref{fitqpbs}, the
specific strangeness content, $s_f/B$ 
along with specific strangeness asymmetry $(\bar s_f-s_f)/B$ 
seen in the hadronic particles emitted.
In the data analysis the requirement that the number of $s$ and $\bar s$ 
quarks in hadrons is equal is in general not enforced.
 We see that in lower portion of 
table  \ref{fitqpbs} that this result is found 
automatically for the symmetric Pb--Pb collision system. However,
a 3.5\,s.d. effect is seen in the asymmetrical S--Au/W/Pb system
Though the  errors which we derive from the experimental data are small, 
there could be in this asymmetric system  a considerable systematic 
experimental error due to data 
 extrapolations made in presence of a significant spectator 
matter component, coupled with theoretical error from 
the varying CM-rapidity. On the 
other hand, the consistency of the Pb--Pb and S--Au/W/Pb
results suggest that this asymmetry is possibly a real effect, 
thus  the unseen balance of strangeness could be hidden in
a residual (strange) quark matter nugget, which is escaping 
detection.  Such strangeletts could in principle form, since in
the hadronization of the S--Au/W/Pb deconfined system the 
hadron phase space is asymmetric, which leads to strangeness
distillation.\cite{Gre87,Raf87,Hei87,Gre91}

One of the important quantitative results of this analysis 
is shown in the bottom section of table~\ref{fitqpbs}: the
 high yield of strangeness per baryon, $s_f/B\simeq 0.7$\,.
We now proceed  to verify if this yield is in agreement  with the 
pedictions made over the  past 20 years. Perhaps more by chance than 
design,  this analysis result is in agreement with the first
calculations of strangeness production employing perturbative QCD,\cite{RM82} 
where the value $N_s/B=n_s/\nu=0.7$ is reached for the plasma temperature of 
300\,MeV as shown there in Fig.\,3. Since, considerably more refined methods 
have been developed,\cite{acta96} and these  are in excellent 
agreement with results of the analysis of
 experimental results. In view of the 
high precision reached in this data 
analysis, we have recomputed the theoretical yield 
taking  for the QCD parameters values generally accepted today: 
$\alpha_{\rm s}(M_Z)=0.118$  and $m_{\rm s}(1GeV)=200$\,MeV, 
correpsonding to $m_{\rm s}(M_Z)=90$\,MeV.

In table \ref{SPRodINput}, we summarize for three collision systems we consider 
S--Au/W/Pb, Ag--Ag, Pb--Pb the 
key input parameters used in computing the result for  $N_s/B$ 
shown below in Fig.\,\ref{figsbrunE}. 
The first entry line gives the central collision particpant numbers for the
three systems considered.  
Next, in table~\ref{SPRodINput}, we see the initial temperature $T_{\rm ch}$ 
which the evaluation of strangeness production  requires as input. 
$T_{\rm ch}$ is the temperature at the  time when light quarks and gluons 
reach equilibrium.  To obtain  this value, we compute the collisional 
pressure and set it equal to 
thermal pressure at the time the fireball begins to expand.\cite{acta96,Let97} 
To do this we need the (momentum, energy) stopping fractions $\eta$ here taken
from NA35/NA49 experimental results,\cite{Alb95} 
(except for interpolation for Ag--Ag,  the dotted line in Fig.\,\ref{figsbrunE}).
The last line in table~\ref{SPRodINput} addresses the 
expansion dynamics we use: we  employ the observed freeze-out
expansion velocity $v_c$ as given in the top section of table~\ref{fitqpbs}. 
We assume that each local volume expands its size scale $R$ at this local velocity,
and we consider the process to be entropy conserving, hence we use $R^3T^3=$Const.
to  obtain the time dependence of local fireball temperature.

\begin{table}[tb]
\caption{\label{SPRodINput}
Input to the strangeness  production computation in QGP.}
\begin{center}
\begin{tabular}{ll|c|c|c}\\
\hline
 &                              &S--Au/W/Pb & Ag--Ag& Pb--Pb  \\
\hline
 participants         & $B$          &  90   &180   &360    \\
 $q,G$ equilibration & $T_{\rm ch}$[MeV] & 260& 280   & 320 \\
 stopping            & $\eta$       &  52\% & 54\% & 57\%  \\
 scale expansion velocity &$v_c$       &0.49$c$   &0.52$c$  &0.54$c$\\
\hline
\end{tabular}
\end{center}
\end{table}

\begin{figure}[tb]
\vspace*{1.9cm}
\centerline{\hspace*{-.5cm}
\psfig{width=10cm,figure=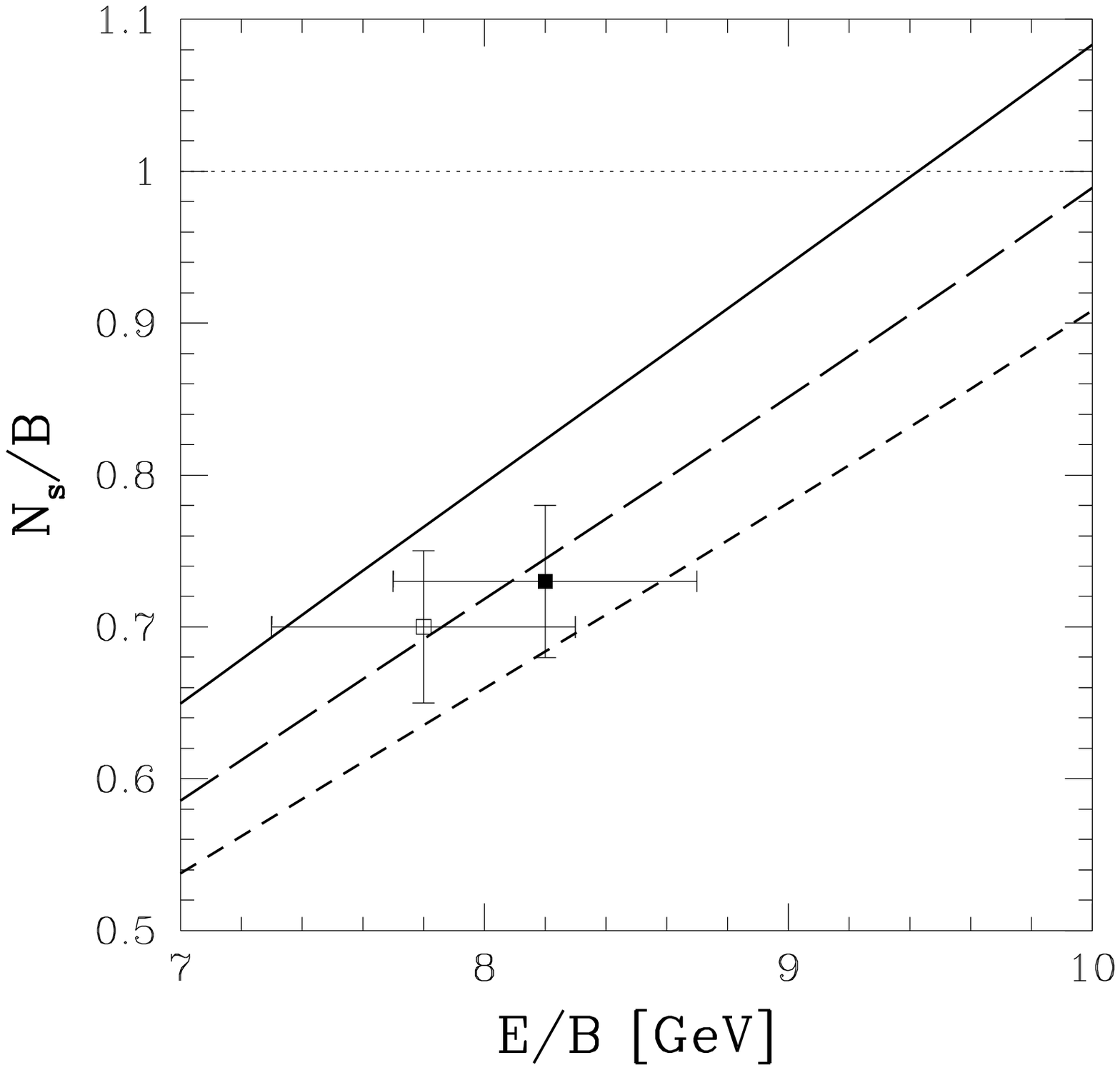}
}
\vspace*{0.2cm}
\fcaption{ 
QGP fireball specific per baryon strangeness abundance  
as function of $E/B$ energy per baryon
in the fireball, for running  
$\alpha_{\rm s}(M_Z)=0.118$ and $m_{\rm s}(1GeV)=200$\,MeV,
correpsonding to $m_{\rm s}(M_Z)=90$\,MeV.
Solid lines: for Pb--Pb, stopping 57\% and 360 particpants, 
long-dashed lines: for Ag--Ag, stopping 54\% and 180 participants,
short-dashed lines: for S--Au/W/Pb stopping 52\% 
and 90 participants in QGP fireball. The solid square is the
result of an analysis for S-Au/W/Pb 200$A$ GeV reaction system, 
and open square for Pb--Pb at 158$A$ GeV, as shown in the
lower section of table~\ref{fitqpbs}. 
\label{figsbrunE}}
\end{figure}

We obbtain  the result for $N_s/B$
shown in Fig.\,\ref{figsbrunE}, as function of the specific energy available 
in the fireball $E/B$, for the three collision systems S-Au/W/Pb (short-dashed
line), Ag--Ag (long dashed) and Pb--Pb (solid line). 
Since we compute the intial temperature from the collision energy 
our approach allows
us to extrapolate as function of $E/B$, assuming that the 
stopping fraction for the collisional pressure is known. When we keep the 
stopping fraction constant and as given 
at the 160--200$A$ GeV collision energy, 
we find the results shown in Fig.\,\ref{figsbrunE}.  
However, a constant stopping
underestimates the intial temperature at lower collision energy, where 
we would expect higher stopping, and it 
overestimates the initial temperature at 
higher collision energies, where 
we would expect smaller stopping, thus we believe that
the slope of the result we present in  Fig.\,\ref{figsbrunE} is too steep.
We will be  able to improve on this result after the  behavior of stopping 
as function of collision energy has been understood.

In Fig.\,\ref{figsbrunE}, the solid square is the
result of the analysis for S-Au/W/Pb system, and open square for Pb--Pb 
as shown in the lower section of table~\ref{fitqpbs}. 
We note that the reason that the available energy $E/B$ in 
the fireball is the dominant parameter controlling strangeness yield is 
the cancellation of effect of higher initial temperature in the larger,
more stopping systems, by the  faster explosion of such a system, 
which leaves less time for strangeness production. We note that 
even though we did not analyze here the S--S system, for which case
we would need to adapt the method to allow significant longitudonal flow, 
it is understood that the available fireball energy and strangeness content per
baryon is higher in S--S 200$A$ GeV interactions,\cite{SGR94}
consistent with the results shown in Fig.\,\ref{figsbrunE}.

This high strangeness yield corresponds to 
(above) equilibrium abundance phase space occupancy 
in hadronization. In the top  section of table~\ref{fitqpbs}, 
the ratio $\gamma_{s}/\gamma_{q}\simeq 0.8$,
which corresponds (approximately)  to the parameter $\gamma_{s}$ when
$\gamma_{q}=1$ has been  assumed.  We observe that  $\gamma_{s}^{\rm Pb}>1$.
This strangeness over-saturation effect could arise from the effect 
of gluon fragmentation combined with early chemical equilibration
in the QGP, $\gamma_{s}(t<t_f)\simeq 1$. The ensuing rapid expansion
preserves this high strangeness yield, and thus we find the result
$\gamma_{s}>1$\,, as we reported in Ref.\,\cite{acta96}.
This high phase space occupancy is
one of the requirements for  the enhancement of multi-strange (anti)baryon
production, which is an important hadronic signal of
QGP phenomena.\cite{firstS} 

\begin{figure}[tb]
\vspace*{1.3cm}
\centerline{ 
\psfig{width=6.8cm,figure=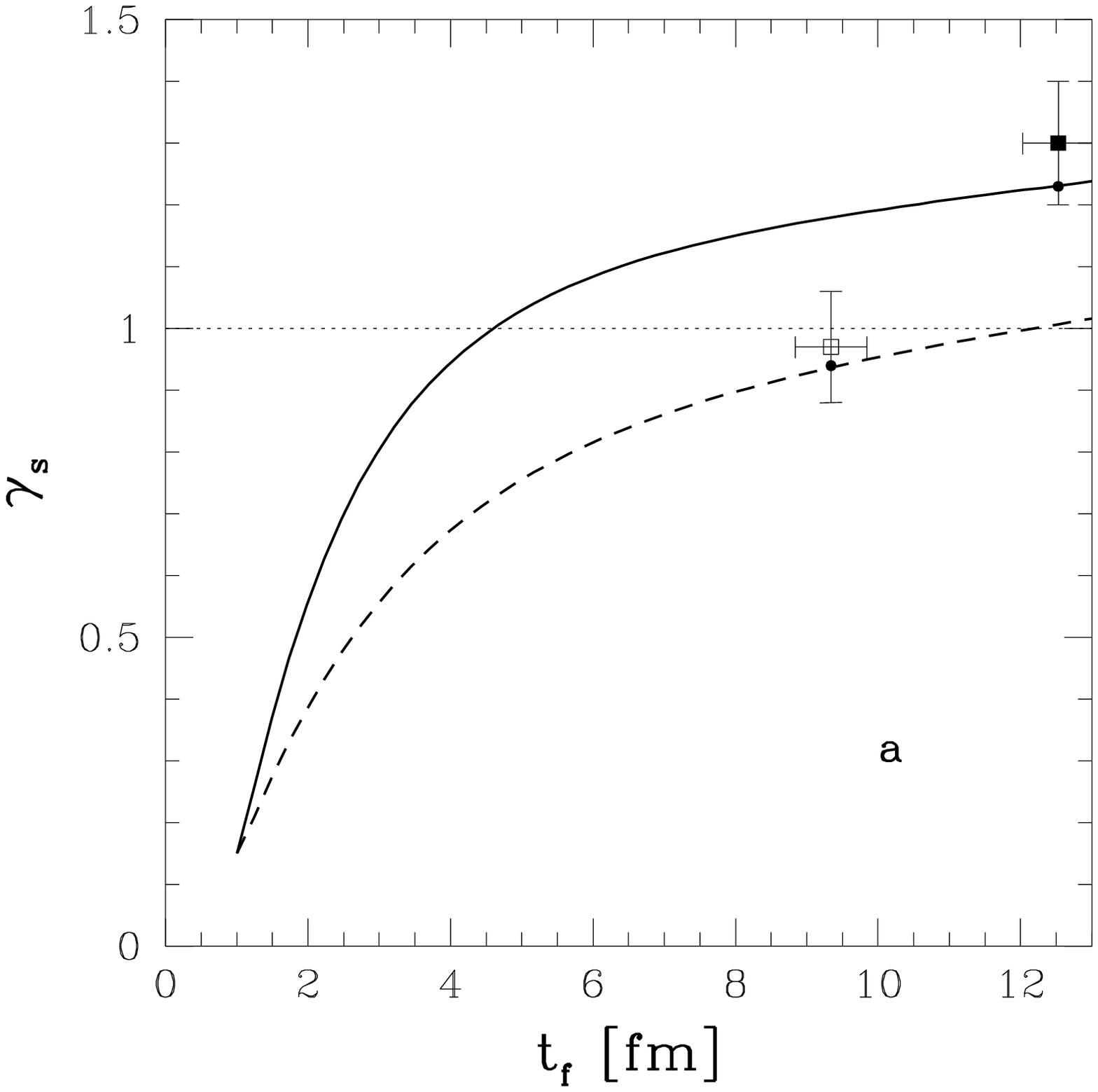}\hspace*{-0.5cm}
\psfig{width=6.8cm,figure=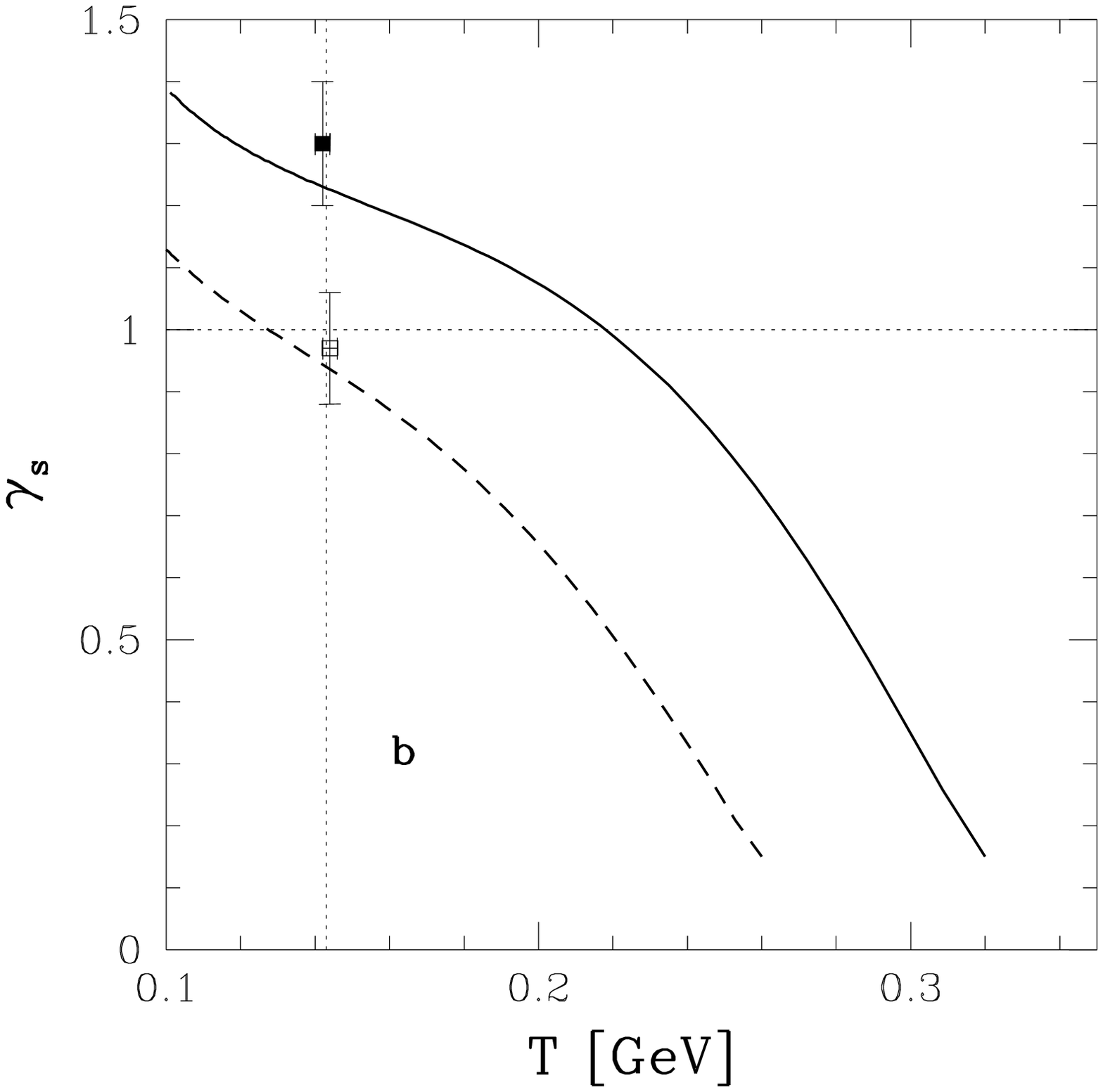}}
\fcaption{ 
Evolution of strangeness phase space occupancy $\gamma_s$, 
{\bf a)}  as function of time $t$ (on left) and 
{\bf b)} as function of temperature $T$ (on right) for the 160--200$A$\ GeV collision systems. 
For further explanation see legend to figure 
\ref{figsbrunE} and text.
\label{gammaSPB}} 
\end{figure}

We compare this result of data analysis, in quantitative 
manner, with the theoretical computation of $\gamma_s$ which is 
easily obtained from the above study of total strangeness production,
as we only need to divide the total momentary strangeness yield by the
expected equilibrium abundance, for which we choose to consider ideal
gas of strange quarks with QCD running mass $m_s(\mu=5.5 T)$\,. 
The factor 5.5 converts the value of 
$T$ into the appropriate scale $\mu$ of energy at which 
the kinetic equilibrium distribution is formed, and we 
note that $m_s(T=182\,{\rm MeV})=200$\,MeV. The effect of 
QCD running influences the agreement between theory and experiment at the 
level of 10--15\%.  The result is shown 
in figure \ref{gammaSPB}, right
as function of time $t$ for the 160--200$A$\ GeV collision systems
and left as function of temperature $T$.
Horizontal dotted line refers to equilibrium phase space occupancy, and the
vertical line indicates expected freeze-out condition at $T_f=143$\,MeV.

Solid dots in figure \ref{gammaSPB}{\bf a)} show where this freeze-out 
temperature occurs as function of time $t$. The analysis point 
uncertainty in freeze-out time is obtained assuming that,
in isentropic evolution, size scale $R$ and temperature satisfy $RT=$ Const., 
and thus with $v_c=dR/dt\vert_f$ we find:
\begin{equation}
\Delta t= -\frac{R_f}{v_cT}\Delta T\,.
\end{equation}
The chemical freeze-out occurs at about 10\,fm/$c$ and 13\,fm/$c$,
after onset of the collision, allowing for about
1\,fm/$c$ initial time $\tau_{\rm ch}$ for the two systems S--Au/W/Pb and Pb--Pb 
respectively. 
Considering that the
expansion velocity has been $0.5c$ and $0.64c$ respectively we
obtain an  estimate of the chemical freeze-out radius 
$ R_f^{\rm S}\simeq 5$fm and $ R_f^{\rm Pb}\simeq 8$fm, the latter value is
in excellent agreement with the discussion of the Coulomb effect presented in 
section~{\bf 2.2}.

\subsection{The Omega riddle}
\noindent
The QGP formation and sudden disintegration model we have 
described above has natural limitations. 
When we attempt to describe within this approach most rarely
produced particles, there is the potential for under-prediction
of experimental results, which could receive contributions
from other more effective production  mechanisms. In this context, 
the most rarely produced hadron is the triply strange 
$\Omega(sss)$ and  $\overline\Omega(\bar s\bar s\bar s)$ which are 
the heaviest stable hadrons, $M_\Omega=1672$\,MeV. The phase space for 
$\Omega$ is more than 10 times smaller than that for  $\Xi$  at the 
conditions of chemical  freeze-out we have obtained. 
$\Omega$ and $\overline\Omega$ production pattern can thus  be 
altered by  processes not implemented in the 
one stage fireball model used to analyze the data.  

When we attempted to describe along with the other hadrons the
yields of $\Omega$ and $\overline\Omega$ within the single stage 
freeze-out model, we indeed have discovered considerable loss 
of physical significance.\cite{LRPb99} Already for the S--Au/W/Pb case, 
we have found that a more reliable description of the data arises if
we did not consider the qualitative
Omega yields available.\cite{LRa99}
For the parameters as  reported, we find in the Pb--Pb reactions
that we under-predict the $\Omega$ and $\overline\Omega$ yields by about factor 2.
The experimental results are  shown in the first three
columns of table \ref{OmegaTab} and the theoretical yield computed using the 
Fermi-2000 model with parameters fixed by other particle abundances 
are  shown in columns 4 and 5: we see that the presence of radial flow ($v=v_c$)
has a minimal impact on the relative yields, compared to the case without 
radial flow ($v=0$).
To put this result into proper perspective, consider that
 we find within the sudden hadronization
of QGP with uncorrelated strange quarks in the deconfined phase an
enhancement of $\Omega$ and $\overline\Omega$ yields `only' by factor 10
as compared to what is expected from extrapolation of p--A 
reactions. However, the experiment reports an enhancement by factor 15--20. 
Such a `failure' is in fact confirming the early expectations that 
$\Omega$ and $\overline\Omega$ yields are the best signature of 
deconfinement, considering the possibility 
of strange quark clustering.\cite{firstS} In fact it is a bit surprising
how well this early prediction works, and this requires further
study to understand more precisely  what exactly this means. 

\begin{table}[tb]
\caption{\label{OmegaTab}
 Pb--Pb 158$A$ GeV particle ratios involving 
$\Omega$ and  $\overline\Omega$, compared to theoretical
expectations for $T_f=143$\,MeV. $^*$ reminds us that statistical parameters 
 are fixed by other particle yields. Last column presents results
allowing a 11\% $\Xi$ and $\overline\Xi$ shadow, see text for details.}
\begin{center}
\begin{tabular}{lcc|ll|l}\\
\hline
Ratios & Ref. &  Exp. Data  &  $v=0$ & $v_c$ & 11 \%  \\
\hline
${\Omega}/{\Xi}$      &\cite{Kra98} &0.192 $\pm$ 0.024                   &0.078$^*$&0.077$^*$ & 0.186\\
${\overline{\Omega}}/{\overline{\Xi}}$  &\cite{Hol97} &0.27 $\pm$ 0.06   &0.17$^*$ &0.18$^*$ & 0.28\\
${\overline{\Omega}}/{\Omega}$  &\cite{Kra98} &0.38 $\pm$ 0.10           &0.57$^*$ &0.60$^*$ & 0.38\\
$(\Omega+\overline{\Omega})\over(\Xi+\bar{\Xi})$&\cite{Hol97}&0.20 $\pm$ 0.03
                                                                         &0.10$^*$ &0.10$^*$ & 0.206\\
\hline
\end{tabular}
\end{center}
\end{table}

Several groups have noted, 
studying the microscopic  evolution of  $\Omega$ and $\overline\Omega$, 
that  due to low reaction cross section they decouple from hadron 
background somewhat sooner than all the other hadrons.\cite{HSX98,Bas99}
An early chemical freeze-out would impact statistical yields 
of $\Omega$ and $\overline\Omega$ greatly.  To augment the 
$\Omega$ and $\overline\Omega$ yields by factor $k$, it is sufficient 
to take an incrementally $\delta T$ higher freeze-out temperature, as determined 
from study of the $\Omega$ phase space: 
\begin{equation}
\delta T\simeq T \frac{\ln k}{M_\Omega/T}\,.
\end{equation}
Thus in order to increase the yields  by a factor 2 the
$\Omega$ and $\overline\Omega$ freeze-out would need to occur
at $T_\Omega=150$\,MeV rather than at  $T_f=143$\,MeV.
Since the temperature drops as the explosion of the fireball develops,
this higher freeze-out temperature means an earlier in time 
freeze-out. 

Even if the required staging in time of hadron production  is 
apparently small, a consistent picture requires fine-tuning
and it seems unnatural, considering that all the other particles 
are perfectly consistent with just one sudden freeze-out condition.
Pursuing other alternatives, we note that  $\Omega$ and $\overline\Omega$ 
enhancement is caused by strangeness 
pre-clustering in the deconfined phase which would enhance multistrange 
hadrons, but most prominently and noticeable enhance the phase space 
suppressed $\Omega$ and $\overline\Omega$. 
In this context, it is interesting to note that the missing yield is not symmetric: 
as seen in table \ref{OmegaTab} we miss in relative terms more $\Omega$ than 
$\overline\Omega$ . Interestingly, the missing yield is exactly 
proportional to the yield of $\Xi$ and $\overline\Xi$ and the best description
of all particle yields, including all $\Omega$ and $\overline\Omega$ is arrived at 
describing what is missing as proportional  (11 \%)
to the $\Xi$ and $\overline\Xi$
yield, this is shown in the last column `11 \%' of table \ref{OmegaTab}. It
is now easy to propose a model that would lead just to this result: 
there are colored di-strange quarks clusters at hadronization
and when their color strings break $\Xi$ and $\Omega$ are produced. This imprints 
a `shadow' of $\Xi$ and $\overline\Xi$ in the $\Omega$ and $\overline\Omega$-abundance.
While this works for $\Omega$ and $\overline\Omega$, 
we find that this mechanism is not compatible with the other 
particle abundances, in other words a similar 
`shadow' of $\Lambda$ and $\overline\Lambda$
in the $\Xi$ and $\overline\Xi$ channel seems  unacceptable. Thus this mechanism 
would work only if pairing of strange quarks would be significant near to phase 
transition. Current models of `color super conductivity' 
support such a clustering  mechanism for additional $\Omega$ and $\overline\Omega$
enhancement, though detailed studies are still in progress.\cite{supercolor}

 We have also explored the 
possibility that unknown $\Omega^*$ and $\overline{\Omega^*}$
resonances contribute to the $\Omega$ and $\overline\Omega$ yield, but
we were not able to find a good set of parameters for these 
hypothetical resonances. Moreover, 
this hypothesis implies a baryon--antibaryon symmetric contribution
in the sense that both  $\Omega$ and $\overline\Omega$ yields
are multiplied by the same factor. However,  the missing yield is clearly 
also baryon--antibaryon asymmetric --- thus despite several ad-hoc
parameters the model description remains poor. 

We note that earlier statistical descriptions of 
$\Omega$ and $\overline\Omega$ yields have not been sensitive to the
problems we described.\cite{Bec98,LRT98} In fact as
long as the parameter $\gamma_q$ is not considered, it is not possible
to describe the experimental data at the level of precision that
would allow recognition of the  $\Omega$ and $\overline\Omega$ yield
as a problem for the statistical Fermi phase space model.

\section{Kinetic Strangeness Production}
\label{sprodsec}
\noindent
In some computational details, the methods to describe strangeness
production differ.\cite{Bir93,Won96,Sri97,RL99}
This leads to different expectations regarding chemical 
equilibration of quark flavor at RHIC energies, with some authors
finding marginal at best chemical equilibration.
We therefore develop in more detail the computational
 approach which is consistent with the SPS-energy scale 
results discussed in previous sections.\cite{RL99}
One important difference to the earlier work is that the 
two loop level running of  QCD parameters for both 
coupling strength $\alpha_s$ and strange quark mass $m_s$ 
is used.  $\alpha_{M_Z}=0.118$ is assumed as determined 
 at the $\mu=M_{Z^0}$ energy scale. Another improvement is that
an entropy conserving explosive flow of matter is 
incorporated directly into the dynamical equations describing
the evolution of strangeness phase space occupancy. 
This approach is entailing significant cancellations 
in the dynamical equations and the only model dependence on
matter flow which remains is the 
relationship between the local temperature and local proper time.
 In consequence, 
a relatively simple and physically transparent model for the 
evolution of the phase space occupancy $\gamma_{s}$ 
of strange quarks in the expanding QGP can be studied. 

We use two assumptions of relevance for the
results we obtain:\\
$\bullet$ the kinetic (momentum distribution) 
equilibrium is reached faster than the chemical (abundance) 
equilibrium\cite{Shu92,Alam94};\\
$\bullet$ gluons  equilibrate chemically significantly 
faster than strangeness.\cite{Won97}\\
The first assumption  allows us to 
study only the chemical abundances, rather than
the full momentum distribution, which simplifies greatly
the structure of the master equations; the second  assumption 
allows us to consider the evolution of the strangeness population
only  after an initial time $\tau_{\rm ch}$ period 
has passed: $\tau_{\rm ch}$ 
is the time required for the development to near chemical
equilibrium of the  gluon population, and the corresponding 
temperature $T_{\rm ch}$ is the initial condition we need
to compute the evolution of strangeness. Aside of $T_{\rm ch}$, 
the strange quark mass $m_{s}$ introduces the greatest uncertainty 
that enters strangeness yield calculations based on resumed
perturbative QCD rates.\cite{budap} The  overpopulation of
the strangeness phase space, seen before in 
section {\bf 3} in SPS data, 
arises in particular for $T_{\rm ch}>250$\,MeV and 
values of strange quark mass  $m_{s}$(1GeV)$\simeq 200\pm20$\,MeV.

In view of these assumptions the phase space  distribution  $f_{s}$  
can be  characterized by a local 
temperature $T(\vec x,t)$ of a (Boltzmann) equilibrium distribution  
$f_{s}^\infty$\,, with  normalization set 
by a phase space occupancy factor:
\begin{equation}\label{gamdef2}
f_{s}(\vec p,\vec x; t))\simeq \gamma_{s}(T) 
   f_{s}^\infty(\vec p;T)\,.
\end{equation}
Eq.\,(\ref{gamdef2})  invokes in the momentum independence of 
$\gamma_{s}$ the  first assumption. More generally,
the factor $\gamma_i,\, i=g,q,s,c$, allows the local density 
of gluons, light quarks, strange quarks and charmed quarks,
respectively to evolve 
independently of the  local momentum shape.  With variables 
$(t,\vec x)$ referring to an observer in the laboratory 
frame, the chemical evolution can be described by the strange
quark  current non-conservation arising from strange quark 
pair production described by a Boltzmann collision term:
\begin{eqnarray}
\partial_\mu j^\mu_{s}\equiv  {\partial \rho_{s}\over \partial t} +
   \frac{ \partial \vec v \rho_{s}}{ \partial \vec x}
&\!=& \!\frac12
\rho_g^2(t)\,\langle\sigma v\rangle_T^{gg\to s\bar s}\nonumber\\  
&+&\!\rho_{q}(t)\rho_{\bar q}(t)
\langle\sigma v\rangle_T^{q\bar q\to s\bar s}\!\! -\!
\rho_{s}(t)\,\rho_{\bar{\rm s}}(t)\,
\langle\sigma v\rangle_T^{s\bar s\to gg,q\bar q}\!.
\label{drho/dt1}
\end{eqnarray}
The factor 1/2  avoids double counting of gluon pairs.
The implicit sums over spin, color and any other 
discreet quantum numbers are combined in the particle density 
 $\rho=\sum_{s,c,\ldots}\int d^3p\,f$, and  we have also 
introduced the  momentum averaged production/annihilation
thermal reactivities (also called `rate coefficients'):
\begin{equation}
\langle\sigma v_{\rm rel}\rangle_T\equiv
\frac{\int d^3p_1\int d^3p_2 \sigma_{12} v_{12}f(\vec p_1,T)f(\vec p_2,T)}
{\int d^3p_1\int d^3p_2 f(\vec p_1,T)f(\vec p_2,T)}\,.
\end{equation}
$f(\vec p_i,T)$ are the relativistic Boltzmann/J\"uttner
distributions of two colliding particles of momentum $p_i$, $i=1,2$. 

The  current conservation used above in the laboratory
`Eulerian' formulation  can also be written with reference to the 
individual particle dynamics in the so called `Lagrangian' 
description: consider
$\rho_{s}$  as the  inverse of the small
volume available to each particle. Such a volume is defined 
in the local frame of reference for which 
the local flow vector vanishes 
$\vec v(\vec x,t)|_{\mbox{\scriptsize local}}=0$.
The considered volume $\delta V_l$ being 
occupied by small number of particles 
$\delta N$ ({\it e.g.}, $\delta N=1$), we have:
  \begin{equation}\label{Nsinf}
\delta N_{s}\equiv \rho_{s} \delta V_l \,.
\end{equation}
 The left hand side (LHS) 
of Eq.\,(\ref{drho/dt1}) can be now written as: 
\begin{equation}\label{LagCor}
{\partial \rho_{s}\over \partial t} +
   \frac{ \partial \vec v \rho_{s}}{ \partial \vec x}\equiv
       \frac{1}{\delta V_l}\frac{d\delta N_{s}}{dt}= 
\frac{d\rho_{s}}{dt}+\rho_{s}\frac1{\delta V_l}\frac{d\delta V_l}{dt}\,.
\end{equation}
Since $\delta N$ and $\delta V_l dt$ are L(orentz)-invariant, 
the actual choice of the frame of reference in which the 
right hand side (RHS) of Eq.\,(\ref{LagCor}) is 
studied is irrelevant and we drop henceforth the subscript $l$.

We can further adapt Eq.\,(\ref{LagCor}) to the dynamics we pursue:
we introduce  
$\rho_{s}^\infty(T)$ as the (local) chemical equilibrium  abundance 
of strange quarks, thus $\rho=\gamma_{s}\rho_{s}^\infty$.
We evaluate the equilibrium abundance  
$\delta N_{s}^\infty=\delta V\rho_{s}^\infty(T)$
 integrating the Boltzmann distribution:
\begin{equation}\label{Nsinfty}
\delta N_{s}^\infty=[\delta VT^3] {3\over\pi^2} \,z^2K_2(z)\,,
\quad z={m_{s}\over T}\,.
\end{equation}
We will below use: 
$d[z^\nu K_\nu(z)]/dz=-z^\nu K_{\nu-1}$,
where $K_\nu$ is the modified Bessel function of order $\nu$.
The first factor on the RHS
 in Eq.\,(\ref{Nsinfty}) is a constant in time should the
evolution of matter after the
initial pre-thermal time period $\tau_0$ 
be entropy conserving,\cite{Bjo83} and thus 
 $\delta VT^3=\delta V_0T^3_0=$ Const.\,.
We now substitute in Eq.\,(\ref{LagCor}) 
and obtain
\begin{equation}\label{lochem}
{\partial \rho_{s}\over \partial t} +
   \frac{ \partial \vec v \rho_{s}}{ \partial \vec x}=  
\dot T\rho_{s}^\infty\left({{d\gamma_{s}}\over{dT}}+
\frac{\gamma_{s}}{T}z\frac{K_1(z)}{K_2(z)}\right)\,,
\end{equation}
where $\dot T=dT/dt$. Note that, in Eq.\,(\ref{lochem}), 
only  a part of  the usual flow-dilution term
is left, since we implemented the adiabatic volume expansion,
and study the evolution of
the phase space occupancy in lieu of particle density.
The dynamics of the local temperature is the only quantity we need
to model.

We now return to study the collision terms seen on
the RHS of Eq.\,(\ref{drho/dt1}). 
A related  quantity is the (L-invariant)
production  rate $A^{12\to 34}$ of particles 
per unit time and space, defined usually 
with respect to chemically equilibrated distributions: 
\begin{equation}\label{prodgen}
A^{12\to 34}\equiv\frac1{1+\delta_{1,2}} \rho_1^\infty\rho_2^\infty 
             \langle \sigma_{s} v_{12}\rangle_T^{12\to 34}  \,.
\end{equation}
The factor $1/(1+\delta_{1,2})$ is introduced to compensate 
double-counting of identical particle pairs. 
In terms of the L-invariant $A$\,,
Eq.\,(\ref{drho/dt1}) takes the form:
\begin{eqnarray}
&&\dot T\rho_{s}^\infty\left({{d\gamma_{s}}\over{dT}}+
\frac{\gamma_{s}}{T}z\frac{K_1(z)}{K_2(z)}\right)
=
\gamma_g^2(\tau)A^{gg\to s\bar s} +\nonumber\\
&&\hspace{2cm}+\gamma_{q}(\tau)\gamma_{\bar q}(\tau)A^{q\bar q\to s\bar s} 
\!-\gamma_{s}(\tau)\gamma_{\bar s}(\tau)(A^{s\bar s\to gg}
\!+A^{s\bar s\to q\bar q}).
\label{rho-gam}
\end{eqnarray}
Only weak interactions convert quark flavors, thus, 
on  hadronic time scale, we have 
$\gamma_{s,q}(\tau)=\gamma_{{\bar s},{\bar q}}(\tau)$. Moreover, 
detailed balance, arising from the time reversal symmetry of the
microscopic reactions, assures that the invariant rates
for forward/backward reactions are the same, specifically
\begin{equation}\label{det-bal}
A^{12\to 34}=A^{34\to 12},
\end{equation}
and thus:
\begin{eqnarray}
\dot T\rho_{s}^\infty\left({{d\gamma_{s}}\over{dT}}+
\frac{\gamma_{s}}{T}z\frac{K_1(z)}{K_2(z)}\right)
&=&
\gamma_g^2(\tau)A^{gg\to s\bar s}
    \left[1-\frac{\gamma_{s}^2(\tau)}{\gamma_g^2(\tau)}\right] \nonumber\\
&&
+\gamma_{q}^2(\tau)A^{q\bar q\to s\bar s}
    \left[1-\frac{\gamma_{s}^2(\tau)}{\gamma_{q}^2(\tau)}\right]\,.
\label{rho-bal}
\end{eqnarray}
When all $\gamma_i\to 1$, the Boltzmann collision term vanishes, we have
reached equilibrium.

As discussed, the gluon chemical equilibrium
is thought to be reached at high temperatures well before the strangeness
equilibrates chemically, and thus we assume this in what follows, and 
the initial conditions we will study refer to the time at which gluons
are chemically equilibrated. Setting $\lambda_g=1$ (and without a
significant further consequence for what follows, 
since gluons dominate the production rate, also $\lambda_{q}=1$),
we obtain after a straightforward manipulation the dynamical 
equation describing the evolution of the local phase space occupancy
of strangeness:
\begin{equation}\label{dgdtf}
2\tau_{s}\dot T\left({{d\gamma_{s}}\over{dT}}+
\frac{\gamma_{s}}{T}z\frac{K_1(z)}{K_2(z)}\right)
=1-\gamma_{s}^2\,.
\end{equation}
Here, we  defined the relaxation time  $\tau_{s}$  of 
chemical (strangeness) equilibration as the 
ratio of the equilibrium density that
is being approached, with the rate at which this occurs:
\begin{equation}\label{tauss}
\tau_{s}\equiv
{1\over 2}{\rho_{s}^\infty\over{
(A^{gg\to s\bar s}+A^{q\bar q\to s\bar s}+\ldots)}}\,.
\end{equation}
The factor 1/2 is introduced by convention in order for the quantity
$\tau_{s}$ to describe the exponential approach to equilibrium.

Eq.\,(\ref{dgdtf})  is the final analytical result describing 
the  evolution of phase space  occupancy.
Since one generally  expects that $\gamma_{s}\to 1$ in a monotonic 
fashion as function of time, it is important to appreciate that this 
equation allows the range  $\gamma_{s}>1$:
when $T$ drops below  $m_{s}$, and $1/\tau_{s}$ becomes small,
the dilution term (2nd term on LHS)
in Eq.\,(\ref{dgdtf})  dominates the evolution of $\gamma_{s}$\,.
In simple terms,  the high
abundance of strangeness produced at high temperature over-populates
the available phase space at lower temperature, when the equilibration
rate cannot keep up with the expansion cooling.
This behavior of $\gamma_{s}$  has been 
shown for the SPS  conditions allowing explosive  transverse expansion
in subsection~{\bf 3.3}. 
Since we assume that the dynamics of transverse expansion 
of QGP is similar at RHIC as at SPS, 
we  obtain similar behavior for $\gamma_{s}$ in section~{\bf 5} below.

$\tau_{s}(T)$\,, Eq.\,(\ref{tauss}),
has been evaluated using pQCD cross section and
 employing next to leading order
running of both the strange quark mass and QCD-coupling 
constant $\alpha_{s}$.\cite{budap} We believe that this method produces
a result for $\alpha_{s}$ that can be trusted  
down to just below 1\,GeV energy scale which
is here relevant. We employ results obtained with
$\alpha_{s}(M_{Z^0})=0.118$ and 
$m_{s}(\mbox{1GeV})=200$\,MeV; we have shown 
results with $m_{s}(\mbox{1GeV})=220$\,MeV earlier.\cite{RL99}
There is some systematic  uncertainty due to the appearance of the 
strange quark mass as a fixed rather than running value
in both, the chemical equilibrium 
density $\rho_{s}^\infty$ in Eq.\,(\ref{tauss}), and in the dilution term
in Eq.\,(\ref{dgdtf}). We use the value 
$m_{s}(\mbox{1\,GeV})$, with the 1\,GeV energy
scale chosen to correspond to typical interaction scale in the QGP
at temperatures under consideration.

\section{Expectations for Strange Hadron Production at RHIC}
\label{rhicsec}
\noindent
We now combine all recent advances in theoretical models of
strangeness production and data interpretation at SPS 
energies with the objective of making reliable predictions for
the RHIC energy range.\cite{RL99} First we address the question
how much strangeness can be expected at RHIC. The
numerical study of Eq.\,(\ref{dgdtf}) becomes possible  
as soon as we define the temporal evolution of the 
temperature for RHIC conditions.   
We expect that a global cylindrical expansion 
should describe the dynamics: aside of the longitudinal flow, we 
allow  the cylinder surface to expand given the internal thermal pressure.
SPS experience suggests that the transverse matter flow will not 
exceed the sound velocity of relativistic matter $v_\bot\simeq c/\sqrt{3}$.
We recall that for a pure longitudinal expansion local entropy density scales
according to $S\propto T^3\propto 1/\tau$.\cite{Bjo83} 
It is likely that the transverse flow of matter 
 will accelerate the drop in entropy density. We thus
 consider the following  temporal evolution function 
of the temperature:
\begin{equation}\label{Toft}
T(\tau)=T_0\left[
\frac{1}{(1+\tau\ 2c/d)(1+\tau\ v_\bot/R_\bot)^2}
\right]^{1/3}\,.
\end{equation}
We take the thickness of the initial collision region at $T_0=0.5$\,GeV
to be\  $d(T_0=0.5)/2=0.75$\,fm,  and the transverse dimension in nearly 
central Au--Au  collisions to be $R_\bot=4.5$\,fm. 
The time at which thermal initial conditions are reached is
assumed to be $\tau_0=1$fm/$c$. When we vary $T_0$, 
the temperature  at which the  gluon equilibrium is reached, we also
scale the longitudinal dimension according to:
\begin{equation}\label{dofT0}
d(T_0)=(0.5\mbox{\,GeV}/T_0)^3 1.5\mbox{\,fm}\,.
\end{equation}
This assures that when comparing the different evolutions of $\gamma_{s}$ we
are looking at an initial system that has the same entropy 
content by adjusting its initial volume $V_0$. 
The reason we vary the initial temperature $T_0$
down to 300 MeV, maintaining the initial entropy content
is to understand how the assumption about the 
chemical equilibrium of gluons, reached by definition at  $T_0$,
impacts strangeness evolution. 
In fact when considering decreasing  $T_0$  (and thus
increasing $V_0$), the thermal production is turned on
at a later time in the history of the collision.

The numerical integration of Eq.\,(\ref{dgdtf}) is started at 
$\tau_0$, and a range of initial temperatures $300\le T_0\le 600$, 
varying in steps of 50 MeV.   The high limit of the temperature
we  explore exceeds somewhat the `hot glue scenario',\cite{Shu92}
while the lower limit of $T_0$ corresponds to the more conservative
estimates of possible initial conditions.\cite{Bjo83}
Since the initial $p$--$p$ collisions also
produce strangeness,  we take as an estimate of initial 
abundance a common initial value
$\gamma_{s}(T_0)=0.15$. The time evolution in 
the plasma phase is followed up to
the break-up of QGP.  This condition we establish
in view of results of the analysis for SPS presented in 
section {\bf 3}. We recall
that SPS-analysis showed that the system dependent 
baryon and antibaryon  $m_\bot$-slopes of particle spectra are
result of differences in collective  flow in the deconfined 
QGP source at freeze-out. In consequence there is universality of
physical properties of hadron chemical  freeze-out between different
SPS systems. This value is nearly applicable to RHIC conditions,
as can be seen extrapolating the 
phase boundary curve to the small baryochemical potentials.
The QGP break-up temperature 
$T_f^{\mbox{\scriptsize SPS}}\simeq 143\pm5$~MeV  
will see just a minor upward change, and we adopt here 
the value $T_f^{\mbox{\scriptsize  RHIC}}\simeq 150\pm5$~MeV.

\begin{figure}[t]
\vspace*{1.4cm}
\centerline{\hspace*{-0.3cm}\psfig{width=7.4cm,figure=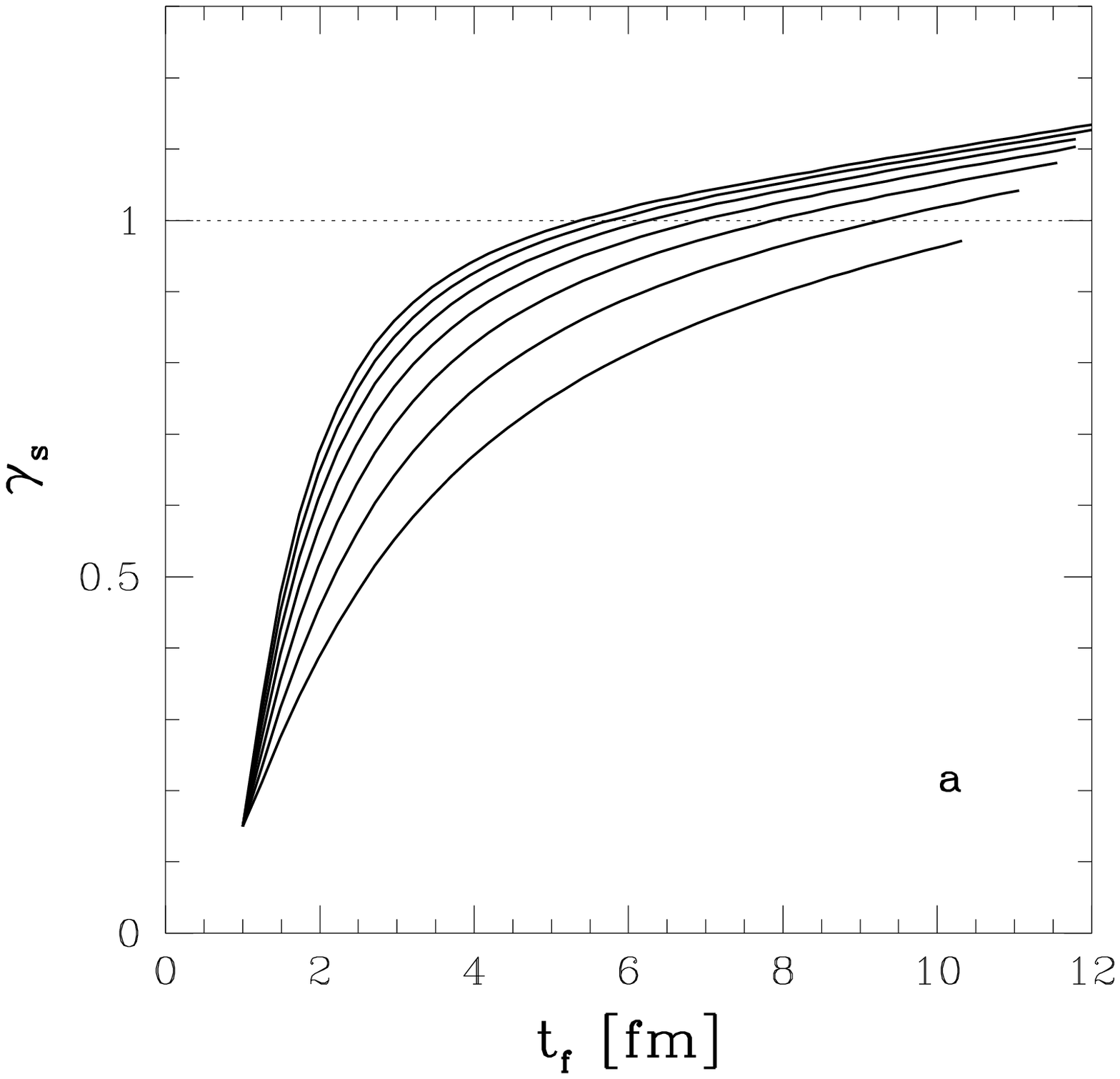}
\hspace*{-1.2cm}\psfig{width=7.4cm,figure=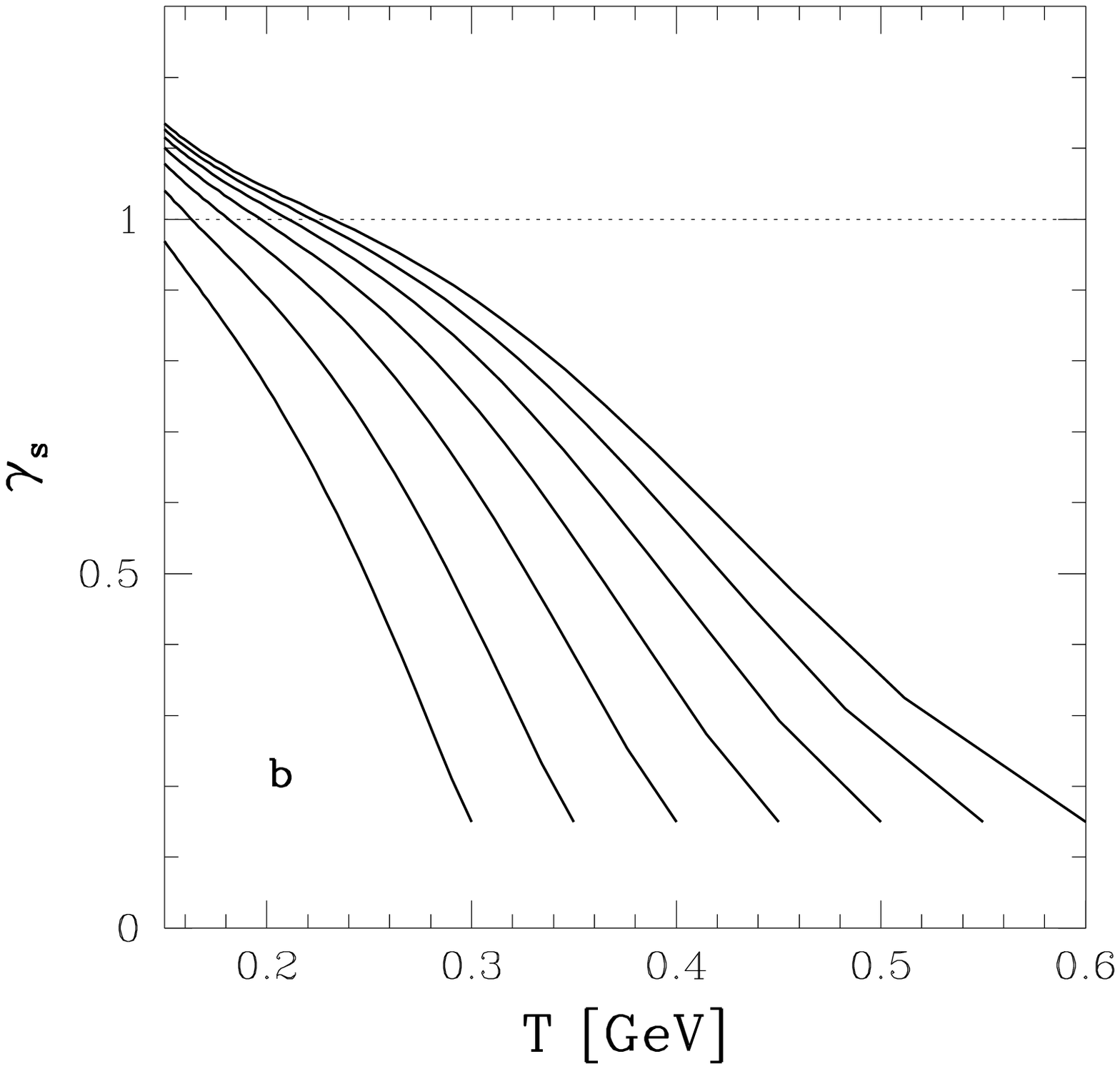}}
\vspace*{-0cm}
\fcaption{ 
Evolution of QGP-phase strangeness phase space occupancy $\gamma_{s}$. 
{\bf a}) as function of  time and, {\bf b}) as function
of temperature for $m_s(1\,\mbox{GeV})=200$\,MeV, see text for details.
\label{figvarsgam}} 
\end{figure}

With the freeze-out condition fixed, 
one would think that the major remaining uncertainty 
comes from the initial gluon equilibration 
temperature $T_0$, and we now study how different values of $T_0$ influence
the final state phase space occupancy. 
We integrate numerically Eq.\,(\ref{dgdtf}) 
and present $\gamma_{s}$ as function of both time $t$ in
Fig.\,\ref{figvarsgam}a, and temperature $T$ in Fig.\,\ref{figvarsgam}b, 
up to the expected QGP breakup at 
$T_f^{\mbox{\scriptsize RHIC}}\simeq 150\pm5$ MeV. 
 We see that: \\
$\bullet$ widely different initial
conditions (with similar initial entropy content) 
lead to rather similar chemical conditions at chemical freeze-out 
of strangeness, \\
$\bullet$  despite
a series of conservative assumptions, we find, not only, that strange\-ness
equilibrates, but indeed that the dilution effect allows an overpopulation of 
the strange quark phase space.
 For a wide range of initial conditions, we obtain a 
narrow band $1.15>\gamma_{s}(T_f)>1$\,.  
We will in the following, taking into account some contribution from hadronization
of gluons in strange/antistrange quarks,  adopt what the value $\gamma_{s}(T_f)=1.25$. 

We now consider how this relatively large value of $\gamma_{s}$,
characteristic for the underlying QGP formation and evolution 
of strangeness, impacts the strange baryon and anti-baryon observable
emerging in hadronization.  Remembering that major changes compared to 
SPS should occur in rapidity 
spectra of mesons, baryons and antibaryons, 
we will apply the same hadronization model that 
worked in the analysis of the SPS data.
This hypothesis can be falsified 
easily, since based and  compared to the 
Pb--Pb 158$A$ GeV results it implies:\\
\indent a) shape identity  
of all RHIC $m_\bot$ and $y$ spectra of antibaryons 
 $\bar p\,,\ \overline\Lambda\,,\ \overline\Xi\,,$ since 
 there is no difference in 
their production mechanism, and the form of the spectra is determined
in a similar way to SPS energy range 
by the local temperature and flow velocity vector; \\
\indent b) the $m_\bot$-inverse-slopes of these antibaryons
 should be very similar to the result obtained at CERN for
  Pb--Pb 158$A$ GeV,
since the expected 3\% increase in the freeze-out temperature 
is accompanied by a comparable 
increase in collective transverse flow.

The abundances of particles produced from QGP  
within the sudden  freeze-out model are controlled by several 
parameters we addressed earlier: the light quark fugacity 
$1<\lambda_{q} <1.1$\,,  value is limited by the expected small
ratio between baryons and mesons (baryon-poor plasma) when the 
energy per baryon is above 100\,GeV, strangeness fugacity 
$ \lambda_{s}\simeq 1$ 
which value for locally neutral plasma assures that 
$\langle s-\bar s\rangle =0$; 
the light quark phase space occupancy 
$\gamma_{q}\simeq 1.5$, overabundance value due to  gluon fragmentation. 
Given these narrow ranges of chemical parameters and 
the freeze-out temperature $T_f=150$ MeV, 
 we compute the expected particle production at break-up. 
In general, we cannot expect that the absolute numbers of particles we
find are correct, as we have not modeled the important effect of flow in
the laboratory frame of reference. However, ratios of
hadrons subject to similar flow effects (compatible hadrons)
can be independent of  the detailed final state dynamics, as 
the results seen at SPS suggest, and
we will look at such ratios more closely.

Taking $\gamma_{q}=1.5\!\!{\tiny\begin{array}{c}+0.10\\-0.25 \end{array}}\!\!$, 
we choose the value of 
$\lambda_{q}$, see the header  of table \ref{table1}, 
for which the  energy  per baryon ($E/B$)
is similar to the collision condition 
(100\,GeV),  which leads to the 
range $\lambda_{q}=1.03\pm0.005$. We  evaluate for 
these  examples aside of $E/B$, the strangeness per baryon 
$s/B$ and entropy per baryon $S/B$ as shown in the top 
section of the table \ref{table1}. We do
not enforce  $\langle s-\bar s\rangle=0$ exactly, but 
since baryon asymmetry is
small, strangeness is  balanced to better than 2\%\, in
the parameter range considered.
In the bottom portion of  table \ref{table1},  we present
the compatible particle abundance ratios,
computed according to the procedure developed 
in { section\bf~2}.  We have given,
aside of the baryon and antibaryon relative yields, also the relative 
kaon yield, which is also well determined within this approach. 
\begin{table}[t]
\tcaption{\label{table1} 
For $\gamma_{s}=1.25,\,\lambda_{s}=1$ and $\gamma_{q}$, $\lambda_{q}$ as shown:
Top portion: strangeness per baryon $s/B$, 
energy per baryon $E/B$[GeV]  and  entropy per baryon $S/B$. Bottom portion:
sample of hadron ratios expected at RHIC.}
\small
\vspace*{-0.2cm}
\begin{center}
\begin{tabular}{llllll}\\
\hline
 $\gamma_{q}$                               & 1.25 & 1.5  &  1.5  &  1.5  & 1.60 \\
$\lambda_{q}$                               & 1.03 & 1.025&  1.03 & 1.035 & 1.03 \\
\hline
$E/B$[{\small GeV}]\ \                        & 117  & 133  &  111  &  95   & 110 \\
$s/B$                                     & 18   & 16   &  13   &  12   & 12 \\
$S/B$                                     & 630  & 698  &  583  & 501   & 571 \\
\hline
$p/{\bar p}$                              & 1.19 & 1.15 & 1.19  &  1.22 & 1.19 \\
$\Lambda/p$                               & 1.74 & 1.47 & 1.47  &  1.45 & 1.35 \\
${\bar\Lambda}/{\bar p}$                  & 1.85 & 1.54 & 1.55  &  1.55 &1.44 \\
${\bar\Lambda}/{\Lambda}$                 & 0.89 & 0.91 &  0.89 &  0.87 & 0.89 \\
${\Xi^-}/{\Lambda}$                         & 0.19 & 0.16&  0.16  &  0.16 & 0.15 \\
${\overline{\Xi^-}}/{\bar\Lambda}$          & 0.20 & 0.17 &  0.17 &  0.17 & 0.16 \\
${\overline{\Xi}}/{\Xi}$                  & 0.94 & 0.95 &  0.94 &  0.93 & 0.94 \\
${\Omega}/{\Xi^-}$                                    & 0.147&0.123 &  0.122&  0.122& 0.115 \\
${\overline{\Omega}}/{\overline{\Xi^-}}$              & 0.156& 0.130&  0.130&  0.131& 0.122 \\
${\overline{\Omega}}/{\Omega}$                      &  1   & 1.   &  1.   &  1.   & 1.   \\
$\Omega+\overline{\Omega}\over\Xi^-+\overline{\Xi^-}$    & 0.15 & 0.13 &  0.13 &  0.13 & 0.12 \\
$\Xi^-+\overline{\Xi^-}\over\Lambda+\bar{\Lambda}$       & 0.19 & 0.16 &  0.16 &  0.16 & 0.15 \\
${K^+}/{K^-}$                                       & 1.05 & 1.04 &  1.05 &  1.06 & 1.05 \\
\hline\vspace*{-0.6cm}
\end{tabular}
\end{center}
\end{table}

The meaning of these results can be better appreciated when
we assume in an example the central  rapidity density
of direct protons is  $dp/dy|_{\mbox{\scriptsize cent.}}=25$. 
In table  \ref{table2},  we present the 
resulting (anti)baryon abundances. 
The  net baryon density $db/dy\simeq 16\pm3$,  there is 
baryon number transparency. We see that (anti)hyperons are 
indeed more abundant than non-strange  (anti)baryons.
Taking into account the disintegration of strange baryons,
we are finding a much greater  number of observed protons 
$dp/dy|_{\mbox{\scriptsize cent.}}^{\mbox{\scriptsize obs.}}
\simeq 65\pm5$ in the central rapidity region. It is important
when quoting results from table  \ref{table2} to recall that:\\
\indent 1) we have chosen arbitrarily the overall
normalization in table  \ref{table2}\,, only particle ratios 
were computed,  and\\ 
\indent 2) the rapidity baryon density
relation to rapidity proton density is a consequence of the assumed 
value of $\lambda_{q}$, which we chose to get 
$E/B\simeq 100$\,GeV per  participant. 

\begin{table}[tb]
\tcaption{\label{table2} $dN/dy|_{\mbox{\scriptsize cent.}}$ 
assuming in this example $dp/dy|_{\mbox{\scriptsize cent.}}=25$ .}
\vspace*{-.2cm} 
\begin{center}
\begin{tabular}{ll|cccccccccc}\\
\hline
$\gamma_{q}$& $\lambda_{q}$  & $b$ &  $p$ & $\bar p$ & $\!\!\Lambda\!\!+\!\!\Sigma^0\!\!$ & $\!\!\overline{\Lambda}\!\!+\!\!\overline{\Sigma}^0\!\!$&$\Sigma^{\pm}$&$\overline{\Sigma}^{\mp}$ & 
$\Xi^{^{\underline{0}}}$ &$\overline{\Xi}^{^{\underline{0}}}$& $\Omega\!=\!\overline\Omega$  \\
\hline
1.25& 1.03 & 17 & 25$^*$& 21 & 44 & 39 & 31 & 27 & 17 & 16 & 1.2  \\ 
1.5 & 1.025 & 13 & 25$^*$& 22 & 36 & 33 & 26 & 23 & 13 & 11 & 0.7 \\ 
1.5& 1.03 & 16 & 25$^*$& 21 & 37 & 33 & 26 & 23 & 12 & 11 & 0.7 \\ 
1.5 & 1.035 & 18 & 25$^*$& 21 & 36 & 32 & 26 & 22 & 11 & 10 & 0.7 \\ 
1.60& 1.03  & 15 & 25$^*$& 21 & 34 & 30 &24 &21 & 10 & 9.6 & 0.6 \\ 
\hline\vspace*{-0.6cm}
\end{tabular}
\end{center}
\end{table}
The most interesting result seen in table  \ref{table2}\,, 
the hyperon-dominance of the baryon 
yields at RHIC, does not depend on detailed 
model hypothesis. We have  explored another set of parameters in
our first and preliminary report on this matter,\cite{prelim} finding this
result. Another interesting property of the hadronizing hot RHIC
matter, as seen in  table \ref{table1}, is that strangeness yield
per participant is expected to be 13--23 times 
greater than seen at present at 
SPS energies, where we have 0.75 strange quark pairs per baryon.
As seen in  table  \ref{table2}, the baryon 
rapidity density is in this examples
similar to the proton rapidity density.

\section{Summary and Conclusions}
\noindent
We believe that this study  of SPS 
strangeness results decisively shows  interesting new 
physics. We see considerable convergence of the results 
around properties of suddenly hadronizing QGP.\cite{LRprl99}
The key results we obtained in the Fermi-2000 model data analysis are: \\
\indent 1) the same hadronization temperature $T$=142--144\,MeV 
for very different collision systems with different hadron spectra;\\
\indent 2) QGP expected results for the source phase space
properties: $\tilde \lambda_{s}=1$  for both
S- and Pb-collisions, implying   $\lambda_{s}^{\rm Pb}\simeq 1.1$\,;\\ 
\indent 3) $\gamma_{s}^{\rm Pb}>1$, indicating that a high strangeness yield was reached before
freeze-out;\\
\indent 4) $\gamma_{q}>1$\, as would be expected from a high entropy phase and the associated
value  $S/B\simeq 43\pm3$\,;\\
\indent 5) the yield of strangeness per baryon $\bar s/B\simeq 0.7$ just as predicted by gluon  
fusion in thermal QGP, a point we studied in detail in section~{\bf 3.3};\\
\indent 6) the transverse expansion velocity for Pb--Pb:
$v_c^{\rm Pb}\le 1/\sqrt{3}$, just below the sound
velocity of quark matter. 

The universality of the physical properties
at chemical freeze-out for S- and Pb-induced
reactions points to a common nature of the
primordial source  of hadronic particles. The difference
in spectra between the two collision systems considered
arises, in this analysis, due to the difference
in the degree of chemical equilibration of light and strange quarks,
expected for systems of differing size and lifespan, and a
difference in the collective surface explosion velocity, 
$v_c^{\rm S}\simeq 0.5\,<\,v_c^{\rm Pb}\simeq 1/\sqrt{3}$\,,
which for larger system  is higher, having  more time to develop. 
Considering how small the experimental WA97 spectral slope errors 
shown in table~\ref{Tetrange2} are presently,  
there is now overwhelming evidence that the production mechanism of 
both $\Lambda$ and $\overline{\Lambda}$ is the same, 
which observation is very probably also true 
for both $\Xi$ and $\overline\Xi$. This symmetry between matter--antimatter 
production is an important cornerstone of the claim that the strange
antibaryon data can only be interpreted in terms of direct emission
from a deconfined and thus matter-antimatter symmetric quark matter. 

We note that the QGP break-up temperature we find, $T_f=143$\,MeV,
corresponds to an energy density 
$\varepsilon=\cal O$(0.5) GeV/fm$^3$.\cite{Kar98} 
Among other interesting results which also verify the consistency
of the experimental data understanding within the Fermi-2000
model, we recall:\\ 
\noindent$\bullet$ the  exact balancing of
strangeness  $\langle \bar s-s\rangle=0$ also in the final hadronic particles
in the symmetric Pb--Pb case;\\ 
\noindent$\bullet$ 
the increase of the baryochemical potential 
$\mu_B^{\rm Pb}={203\pm5} >\mu_B^{\rm S}=178\pm5$\,MeV
as the collision system grows;\\ 
\noindent$\bullet$ 
the energy per baryon near to the value expected if energy and baryon number
deposition in the fireball are similar;\\
\noindent$\bullet$ 
hadronization into pions at  
$\gamma_{q}\to \gamma_{q}^c=e^{m_\pi/2T_f}\simeq 1.6$
seen in Pb--Pb reactions, which is an 
effective way to convert excess of entropy 
in the plasma into hadrons, without need for reheating, or a mixed
phase; the finding of 
the maximum allowable $\gamma_{q}$ is intrinsically consistent with
the notion of an explosively disintegrating QGP phase.

A reassuring feature  of the Pb--Pb analysis related to 
chemical equilibration  has been described in  
{subsection\bf~2.3}:
we find  a pion yield  which  maximizes the entropy density 
of hadronic particles  produced.\cite{LTR00} This 
detailed technical result explains how  sudden  hadronization 
 can occur: in general the deconfined state
with broken color bonds and thus the high entropy density 
has to find an exit into the hadronic world,
maintaining or increasing the total entropy and preferentially 
also the local entropy contained within a small, comoving volume cell.
Our analysis of experimental results suggests that this is 
accomplished by generating an over-saturated
pion phase space, in which the entropy density rises to 
values as high as are believed to occur in QGP at hadronization.
Chemical equilibrium hadronization requires the  formation of  
a mixed plasma-hadron gas phase and is generally believed to
require a relatively long time, followed by kinetic 
reequilibration. In our opinion such a hadronization model is now
 inconsistent with the experimental strange 
baryon and antibaryon data on yields and 
spectra. The reader should note that such
technical differences between different groups about the dynamics
of the evolution of the hadron fireball {\it after} the deconfined 
phase has hadronized, do not impact the primary agreement about the 
deconfined nature of the high density source of hadronic particles. 

We believe that omission to consider chemical non-equilibrium 
in the study of freeze-out conditions employing the analysis of 
spectral shape (flow) and also pion correlation (HBT) effect 
is the source of the difference of results here presented 
with some  other recent work.\cite{HeiF99,WA98freeze} To understand the 
source of this difference it is important 
to realize that there is a considerable 
influence on the shape  of pion spectra by the light quark 
chemical non-equilibrium which the data analysis presented includes: 
the cocktail of resonance decays contributing
to pion spectra is altered, and moreover, there is 
spectral deformation at low $m_\bot$ due to 
pion correlation effects caused by the overpopulated phase space.\cite{LTR00} 
We note that results  presented also differ somewhat 
from the WA98 experiment 
analysis addressing solely $\pi^0$  spectra, and which 
again assumes pion chemical equilibrium.\cite{WA98freeze} 
In consequence,  the $\pi^0$-freeze-out 
conditions as seen  in Ref.\,\cite{WA98freeze}, Table\,1
are different from those determined here. 
 On the other hand, another recent hadron  spectral shape 
analysis,\cite{Sch99} which did not introduce  low $m_\bot$ 
pion spectra into consideration  obtains a chemical freeze-out 
conditions nearly identical to those we discussed. 
While the precise understanding of hadronization condition
is required for a measurement of physical properties 
of QGP including the latent heat, the differences discussed
are of little if any consequence concerning the fundamental 
issue, the question if deconfinement is achieved.

The sudden hadronization of entropy rich QGP 
leads to value $\gamma_{q}\to \gamma_{q}^c$, in
order to  connect the entropy rich deconfined and 
the confined phases more efficiently. The dominant pion contribution to the
entropy density (and pressure) is nearly  twice as high at $\gamma_{q}\simeq 
\gamma_{q}^c$ than at $\gamma_q=1$. Without this phenomenon 
one has to introduce a mechanism that allows the parameter $VT^3$ 
to grow, thus expanding either the volume $V$ due to formation of the
mixed phase or invoking a rise of $T$ in the reheating. 
The range of values for $\gamma_{q}$ is bounded from above by the 
Bose distribution singularity $\gamma_{q}\to \gamma_{q}^c$, 
but a pion condensate is  not formed since it `consumes' energy without
consuming entropy of the primordial high entropy QGP phase. 
 An interesting feature of such a mechanism 
of phase transition which side-steps the need to form 
a mixed phase or reheating is that the chemical non-equilibrium reduces
and potentially eliminates any discontinuity in the phase transition. 
This being the case, experimental searches will not find the critical fluctuations
expected for a discontinuous phase change, even if theory implies a 1st order
phase transition for the statistical equilibrium system. This is in agreement
with the failure  of NA49 experiment to find precritical fluctuations
in event-by-event analysis.\cite{NA49fluc}

The only not fully quantitatively described
 particle yields are $\Omega$ and $\overline\Omega$:
for the parameters we find, the Fermi-2000 model applied to Pb--Pb reactions
under-predicts this smallest of all hadronic abundances  by about factor 2\,.
This means that we expect, within the sudden hadronization
of QGP with uncorrelated strange quarks in the deconfined phase, only an
enhancement of $\Omega$ and $\overline\Omega$ yields by a factor 8-10
as compared to what is expected from extrapolation of p--A 
reactions. Since the experiment reports an enhancement by factor 15--20,
we need to think again.  This `failure' of Fermi-2000 model 
is in fact confirming the early expectation that 
$\Omega$ and $\overline\Omega$ yields are the best signature of 
deconfinement,\cite{firstS} we  just  must  in future address 
the question what exactly this tells us about QGP structure. 
We have argued nearly 20 years ago that strangeness pairing in
the color anti-triplet channel $(ss)_{\bar 3}$ in the
QGP source would enhance $\Omega$ and $\overline\Omega$  
yields,\cite{firstS} a 
point that is of some topical interest today in context
of color superdconductivity studies.\cite{supercolor}

Despite this unexpected additional enhancement, we firmly conclude
in view of all diverse evidence that (multi)strange hadronic 
particles seen  at CERN-SPS
are  emerging from a deconfined QGP phase of hadronic matter
and do not undergo a  re-equilibration after they have been
produced.  This finding has encouraged us to consider within the same
computational scheme the production of strange hadrons at RHIC 
conditions. First, we have shown that one can expect strangeness chemical 
equilibration in nuclear collisions at RHIC if
the deconfined  QGP is formed. There will, as at SPS, be overpopulation 
effect associated with the early strangeness abundance freeze-out
before hadronization. Most importantly for signatures of new physics
at RHIC, we found that strange  
(anti)baryon abundances will be greater than the yields of non-strange
baryons (protons, neutrons). Consequently, the rapidity distributions of
(anti)protons are arising from decays  of (anti)hyperons.

We are not aware that microscopic model studies reported in the literature 
about RHIC conditions which have noted this remarkable 
hyperon dominance result,  see, {\it e.g.}, Ref.\,\cite{Dum99}. 
The reader could wonder 
why is this unusual phenomenon not happening at SPS  
energies described in {section\bf~3}?
At SPS there is still an appreciable 
relative baryon abundance among all hadrons (about 15\%) 
and the strangeness yield is at SPS energies only at a level  similar
to the baryon yield. 
Thus while abundant (anti)hyperon formation begins to set in, 
there are  still  many non-strange (anti)baryons produced. 
With increasing per baryon energy the yield of strange quark pairs
per baryon rises, and at the same time
the relative abundance of baryons among all hadrons diminishes.
As result, at RHIC energies, we have predicted that  hyperons and/or antihyperons
 are the dominant  population fraction among all baryons and/or antibaryons. 
We thus believe that the preponderance of hyperons as the dominant
(anti)baryon population at RHIC energies can be uniquely correlated
with the formation and sudden hadronization of deconfined QGP phase. 

In a nutshell: we find that strangeness and (anti)baryon QGP signatures are 
conclusively proving formation of deconfined quark matter phase
 at SPS energies, and that these signatures of new physics
are much more distinct at higher RHIC energies. \\ 

{\it Acknowledgment:} We thank the editor of {\it Int. J. Mod. Phys.} E, 
Ernest Henley,  and  Keith Dienes  for valuable comments and suggestions. 


\nonumsection{References}

\end{document}

\centerline{\bf INTERNATIONAL JOURNAL OF MODERN PHYSICS E:}
\centerline{\bf REPORTS ON NUCLEAR PHYSICS}
\vspace*{0.4in}
\centerline{GUIDELINES FOR CONTRIBUTORS}
\vspace*{0.2in}
\begin{arabiclist}
\item Manuscripts should be submitted in {\it quadruplicate} to
one of the Editors, the addresses of whom are printed on the
inside front cover.

\item Submission of a manuscript indicates a tacit understanding
that the paper is not actively under consideration for
publication with other journals.

\item Once the paper is accepted authors are assumed to cede
copyrights of the paper over to World Scientific Publishing Co.
If available, please also send the accepted manuscript on a
disk. (Please refer to the Instructions for typesetting
manuscripts using \LaTeX.)

\item All papers will be acknowledged and refereed. They will
not be returned.

\item Submission of manuscripts implies that the manuscripts are
in their final form and will not be returned to the author for
proofreading before publication.

\item There will be no page charge for the journal. 50 free
reprints will be given to the first author of each accepted
paper. Purchase orders for additional reprints can be made on
order forms, which will be sent to authors.

\end{arabiclist}
\end{document}